\documentclass[prb,twocolumn,floatfix,notitlepage,nofootinbib,superscriptaddress,longtable]{revtex4-2}

\usepackage{amsmath, amssymb, amsfonts, bm, esint}
\usepackage{graphicx, epsfig}
\usepackage{psfrag}
\usepackage{overpic}
\usepackage{color}
\usepackage[normalem]{ulem}
\usepackage{lipsum}
\usepackage{verbatim}
\usepackage[figuresright]{rotating}
\usepackage[justification=raggedright,singlelinecheck=false]{caption}
\usepackage[version=4]{mhchem}
\usepackage[svgnames]{xcolor}
\usepackage[colorlinks,linkcolor=blue,anchorcolor=blue,citecolor=blue,urlcolor=blue]{hyperref}

\usepackage{subfig} 

\usepackage{tikz}
\usetikzlibrary{positioning, arrows.meta, shapes.geometric, calc, backgrounds, fit}

\newenvironment{proof}[1][Proof]{\noindent \textbf{#1: }}{\ \rule{0.5em}{0.5em}}

\def\be{\begin{equation}} \def\ee{\end{equation}}
\def\bea{\begin{eqnarray}} \def\eea{\end{eqnarray}}

\renewcommand{\vec}[1]{\mathbf{#1}}

\usepackage{xcolor}

\newcommand{\ket}[1]{| #1 \rangle}
\newcommand{\bra}[1]{\langle #1 |}

\newcommand{\up}{\uparrow}

\definecolor{Qicolor}{RGB}{3, 136, 252}

\DeclareMathOperator{\Tr}{Tr}

\DeclareMathOperator{\diag}{diag}

\begin{document}
\title{Fourier Neural Operators for Time-Periodic Quantum Systems: Learning Floquet Hamiltonians, Observable Dynamics, and Operator Growth}
    
\author{Zihao Qi}
\email[Contact author: ]{zq73@cornell.edu}
\affiliation{Department of Physics, Cornell University, Ithaca, NY 14853, USA.}

\author{Yang Peng}
\email[Contact author: ]{yang.peng@csun.edu}
\affiliation{Department of Physics and Astronomy, California State University, Northridge, Northridge, California 91330, USA}
\affiliation{Institute of Quantum Information and Matter and Department of Physics,California Institute of Technology, Pasadena, CA 91125, USA}

\author{Christopher Earls}
\email[Contact author: ]{earls@cornell.edu}
\affiliation{Center for Applied Mathematics, Cornell University, Ithaca, NY 14853, USA.}

\date{\today}

\begin{abstract}
Time-periodic quantum systems exhibit a rich variety of far-from-equilibrium phenomena and serve as ideal platforms for quantum engineering and control. However, simulating their dynamics with conventional numerical methods remains challenging due to the exponential growth of Hilbert space dimension and rapid spreading of entanglement. In this work, we introduce Fourier Neural Operators (FNO) as an efficient, accurate, and scalable framework for non-equilibrium quantum dynamics. Parameterized in Fourier space, FNO naturally captures temporal correlations and remains minimally dependent on discretization of time. We demonstrate the versatility of FNO through three complementary learning paradigms: reconstructing effective Floquet Hamiltonians, predicting expectation values of local observables, and learning quantum information spreading. For each learning task, FNO achieves remarkable accuracy, while attaining a significant speedup, compared to exact numerical methods. Moreover, FNO possesses capabilities beyond that of conventional methods, such as predicting all local observables from a subset of measurements without information about the Hamiltonian, as well as extrapolating beyond the time window provided by training data, enabling access to observables and operator-spreading dynamics that might be beyond the coherence time. By employing a spatially local basis, we argue that the computational cost of FNOs scales only polynomially with the system size. Our results establish FNO as a versatile and scalable computation framework that integrates numerical simulations and experimental data seamlessly, with direct implications for extracting meaningful physics from measurements by near-term quantum computers.
\end{abstract}

\maketitle

\section{Introduction}
In recent years, manipulating quantum systems through time-dependent driving has emerged as a central paradigm in condensed matter physics and quantum information science. In particular, periodically driven (Floquet) quantum systems exhibit a wide range of far-from-equilibrium phenomena, such as time crystals~\cite{timecrystal1, timecrystal2,timecrystal3}, Floquet topological insulators~\cite{floquettopology,floquettopology2,floquettopology3,floquettopology4, floquettopology5}, and prethermalization~\cite{Else2017, prethermal1,prethermal2}. These phenomena establish Floquet systems as a versatile platform for engineering novel non-equilibrium phases of matter, as well as exploring potential applications to quantum information technologies, such as achieving high-fidelity quantum control~\cite{floquetcontrol2, floquetcontrol3, floquetcontrol4}.

Despite the promise of Floquet quantum systems, numerical simulation of their real-time dynamics remains extremely challenging when interactions are present, even in one dimension. The exponential growth of the Hilbert space dimension quickly renders exact diagonalization infeasible beyond modest system sizes~\cite{expgrowth1,expgrowth2}. At the same time, the rapid growth of entanglement under driving severely limits tensor-network-based time evolution methods, which are otherwise highly effective in static and weakly entangled quantum systems~\cite{EE1, EE_0, dmrg, dmrg2}. As a result, understanding and predicting dynamics in Floquet systems requires new computational paradigms and surrogates.

Recently, machine learning has become a complementary approach for modeling the dynamics of driven quantum systems. By extracting complex patterns hidden within data, machine-learning-based methods have been successful in predicting dynamics of driven and dissipative quantum systems~\cite{DeepLearningCircuit, dual_complexity, random_driving, Hamiltonian_reconstruction, NQP}. In fact, for a wide class of quantum many-body problems, machine-learning algorithms trained on experimental data can have a proven advantage over
traditional classical algorithms that lack such data~\cite{huang_science}.
For periodically driven systems, capturing the inherent time-periodicity is essential for efficient learning and prediction, as well as physically faithful modeling. There is, then, a need for utilizing a machine-learning architecture tailored for Floquet systems that explicitly leverages their time-periodic nature.

Meanwhile, recent experimental progress has provided access to high-fidelity, real-time measurements on driven quantum systems, particularly on Noisy Intermediate-Scale Quantum (NISQ) devices and programmable quantum simulators~\cite{quantumcomputer1,quantumcomputer2}. Those experimental data can be readily obtained from platforms such as trapped ions, ultracold atoms, and superconducting qubit arrays. However, the temporal window of such systems is still limited by decoherence. Developing methods that can learn from short-time
data and reliably extrapolate to longer timescales is therefore of direct importance for probing non-equilibrium phenomena on near-term
quantum hardware.


In this work, we bridge the gap between the access to large-scale, short-time measurements on Floquet systems and the intractability of their numerical simulations. We propose \textit{Fourier Neural Operators} (FNO) as a surrogate for studying Floquet quantum dynamics. The FNO offers two key advantages over traditional neural network methods. First, its architecture naturally aligns with the physics underlying time-periodic systems, through parameterization in Fourier space. Long-range temporal dependencies, which are difficult for conventional neural networks such as recurrent architectures to capture, are transformed into couplings between frequency modes~\cite{FNO}. This enables FNOs to deliver efficient and robust predictions. Second, operator learning methods such as the FNO are generally minimally dependent on the discretization of the underlying domain~\cite{FNO, Kovachki_NeuOp, boulleReview}. In the context of time-periodic systems, this means that an FNO trained on a specific driving frequency and underlying discretization can be used to make accurate predictions beyond its original training conditions, without retraining.


Compared to previous applications of FNOs to quantum dynamics~\cite{spinchain_fno, NQP}, our formulation offers three key advantages.
First, our framework integrates simulation and experimental measurement data together seamlessly, yielding a data-driven quantum propagator that, once trained, can be applied to different system parameters and disorder realizations; even those not present in the training set.
This contrasts with conventional tensor-network time evolution methods~\cite{TEBD1, TEBD2}, which can be accurate for large one-dimensional systems at moderate times, but must be re-run for each parameter set and each disorder instance; making disorder-averaged studies and parameter sweeps expensive.
Second, our formulation supports extrapolation beyond the training window, which is often limited by coherence time in experiments, and enables longer-time prediction from short-time inputs.
Most importantly, by learning dynamics in a local operator basis, rather than an exponentially large Hilbert space representation~\cite{NQP}, our FNO's model size and inference cost scale \textit{polynomially} with system size; making rapid inference across ensembles of driven and disordered Hamiltonians tractable at large system sizes.

Through three complementary learning paradigms, we demonstrate the wide applicability of our framework, which ranges from predicting Floquet Hamiltonians to learning local observables and forecasting operator growth dynamics. Our FNO-based model is not only more efficient and scalable compared to traditional numerical methods, but also has qualitatively distinct yet experimentally relevant capabilities, such as predicting all local observables from a subset of measurements without knowledge of the Hamiltonian and extrapolating dynamics beyond the training window. Our framework provides a scalable bridge between the data produced by near-term quantum devices and the physical predictions needed to understand driven quantum matter.


The rest of this paper is organized as follows. In Section~\ref{sec:background}, we introduce the structure of Fourier Neural Operators, outline the data needed for each learning task, and discuss the concrete physical model. In the following sections, we explore three different yet complementary learning paradigms: in Sec.~\ref{sec:Heff}, we demonstrate that FNO accurately predicts effective Floquet Hamiltonians and thereby stroboscopic evolution of expectation values. Sec.~\ref{sec:observable} discusses learning expectation values of local observables, using both Hamiltonian parameters and partial measurements as inputs. In Sec.~\ref{sec:opgrowth}, we go beyond observable dynamics and demonstrate that the FNO learns how operators spread under Floquet dynamics. The FNO's predictions allow for extrapolating beyond the training window and initial state provided in the data. Finally, we end with a discussion on FNO's scalability and potential directions for future work in Sec.~\ref{sec:discussion} and a conclusion in Sec.~\ref{sec:conclusion}.

\section{Background \label{sec:background}}
\subsection{Review of Fourier Neural Operators \label{sec:fno}}

Operator learning has recently emerged as a powerful framework in the scientific machine learning community, particularly for modeling complex dynamical systems~\cite{op_learning_pde2, op_learning_pde3,op_learning_pde1,Kovachki_NeuOp, deeponet2021}. In contrast to traditional neural networks that map between finite-dimensional Euclidean spaces, operator learning aims to approximate unknown \textit{operators}, i.e., mappings between abstract vector spaces of functions. Neural operators have proven particularly effective for solving partial differential equations (PDEs) and have found applications across various scientific disciplines~\cite{deeponet2021,FNO, Boullé2022,Gin2021_DeepGreen, Li2020_MGNO, Li2023_GINO, Li2020_GraphNO}. 

Formally, consider $d$-dimensional bounded domains $D, D' \subset \mathbb{R}^d$ and separable Banach spaces of vector-valued functions $\mathcal{U, V}$:
\begin{equation}
    \mathcal{U}: D \rightarrow \mathbb{R}^{d_i}; \mathcal{V}:D' \rightarrow \mathbb{R}^{d_o},
    \label{eq:UV}
\end{equation}
where $d_i$ and $d_o$ are the input and output dimensions, respectively. 

The goal of operator learning is to approximate a (generally non-linear) operator between the two function spaces, $\mathcal{A}: \mathcal{U} \rightarrow \mathcal{V}$  given data in the form of pairs of functions $(u_i, v_i)$, where $i$ is an index of the samples, and all $u_i\in \mathcal{U}$ and $v_i \in \mathcal{V}$.  We aim to construct a neural operator $\mathcal{A}'$, such that $\mathcal{A}'(u') \simeq \mathcal{A}(u')$ for any $u' \in \mathcal{U}$. This task is usually achieved by representing $\mathcal{A}'$ as a neural operator parameterized by $\vec{\theta} \in \Theta$, where $\Theta \subset \mathbb{R}^M$ is the space of parameters and $M$ is the number of parameters in the neural operator. Choosing the best approximation $\mathcal{A}'$ amounts to optimizing over the admissible $\theta$. Operator learning thus becomes a supervised learning problem, in which we seek the optimal set of parameters that minimize the following cost function:
\begin{equation}
\min_{\vec{\theta} \in \Theta} \sum_{i} \mathcal{L} \left(\mathcal{A}'(u_i, \theta), v_i \right),
\label{eq:NO_loss}
\end{equation}
where $\mathcal{L}$ is a loss function that measures the difference between the neural operator's predictions $\mathcal{A}'(u_i, \theta)$ and the ground truth $v_i$.

\begin{figure}
    \centering
    \includegraphics[width=\linewidth]{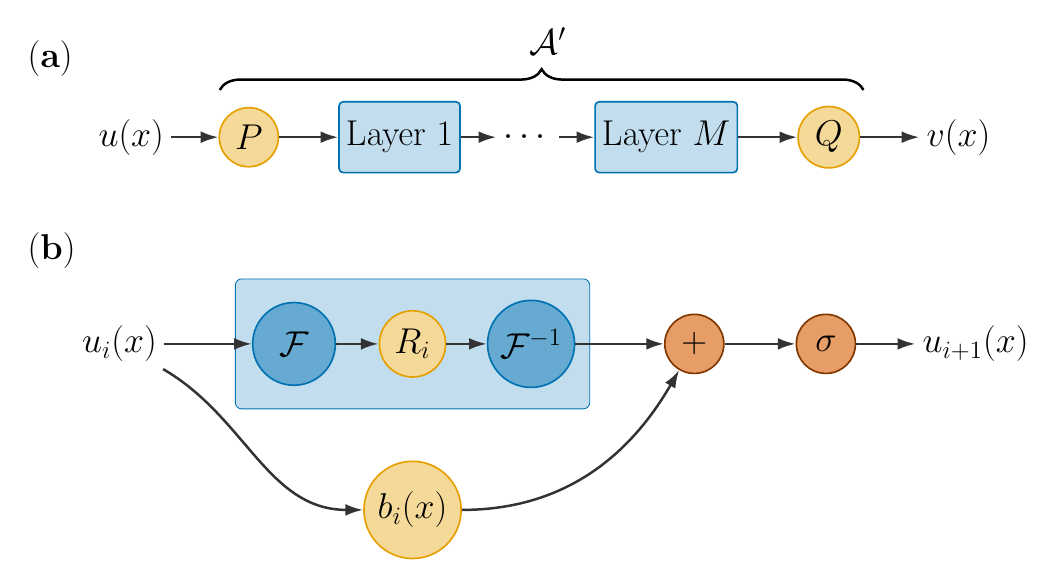}
    \caption{\textbf{(a)} General architecture of a neural operator $\mathcal{A}'$. The input function $u(x) \in \mathcal{U}$ is first lifted to the latent space by a local lifting operator $P$. The $M$ layers of the neural operator iteratively update $u_i(x)$ via an integral kernel described by Eq.~\ref{eq:neural_op}. Finally, a projection operator $Q$ maps the function to the output space $\mathcal{V}$. \textbf{(b)} The architecture of a Fourier layer in FNO. Each Fourier layer performs a spectral convolution over $u_i(x)$, as well as a pointwise linear transformation $b_i(x)$. The two are added element-wise before passed to a nonlinear activation function $\sigma$.}
    \label{fig:FNO}
\end{figure}

The architecture of neural operators generalizes that of neural networks. Consider a typical neural network with $M$ layers. The input vector $v_0$ is iteratively updated as it is passed through layers of the network, $v_0 \rightarrow v_1 \rightarrow ... \rightarrow v_M$. For neural operators, the input is now a function $u(x) \in \mathcal{U}$, where $x \in D$. $u(x)$ is first mapped to a (usually higher-dimensional) latent space via a local lifting operator $P$, $u_0(x) = P(u(x))$. $u_0(x)$ is then recursively updated through layers of the neural operator, $u_0(x) \rightarrow u_1(x) \rightarrow ... \rightarrow u_M(x)$. Finally, a projection layer $Q$ brings $u_M(x)$ back to the output space $\mathcal{V}$. The general architecture of a neural operator $\mathcal{A}'$ is sketched in Fig.~\ref{fig:FNO}(a).

Each layer of the neural operator is composed of integral operators and non-linear functions~\cite{boulleReview}. Generically, the $i$th layer of the neural operator acts on the output of the previous layer (namely $u_i(x)$) as:
\begin{equation}
    u_{i+1}(x) = \sigma \left( \int_D K^i(x, y) u_i(y) dy + b_i(x) u_i(x)\right),
    \label{eq:neural_op}
\end{equation}
where $\sigma(\cdot)$ is a non-linear activation function, $K^i(x, y)$ is the Kernel function, and $b_i(x)$ is a pointwise linear function. $K^i(x, y)$ and $b_i(x)$ are parameterized by the weights $\vec{\theta} \in \Theta$, which are trained similar to ones in standard neural networks.

While the integral operator is linear, neural operators can learn highly non-linear operators due to compositions with the activation functions. In fact, universal approximation theorems have been developed for neural operators~\cite{univ_approx_1, univ_approx_FNO, deeponet2021}, guaranteeing that neural operators can approximate a large class of operators to arbitrary accuracy for sufficiently many data points and model parameters.

Despite neural operators' ability of approximation, there is a major drawback: the integral operator in Eq.~\ref{eq:neural_op} is computationally expensive to evaluate. The \textit{Fourier Neural Operator} (FNO) alleviates this issue through implementations using Fast Fourier Transforms. FNO was first proposed in Ref.~\cite{FNO} as an efficient way to solve parametric PDEs under periodic boundary conditions (PBC). FNO has since been applied to solving various problems in physics, including predicting wavefunction scattering~\cite{scattering_fno}, learning driven-dissipative quantum dynamics~\cite{NQP}, and extrapolating wavefunction and observables in time~\cite{spinchain_fno}.

The key assumption within an FNO is that the kernel function $K^i(x,y)$ is translationally invariant: $K^i(x, y) = K^i(x-y)$. The integral in Eq.~\ref{eq:neural_op} thus becomes a convolution operator.  This allows for parameterizing the Fourier layers in Fourier space directly:
\begin{equation}
    u_{i+1}(x) = \sigma \left( \mathcal{F}^{-1}(R_{i} \cdot \mathcal{F}(u_i)) (x) + b_i(x) \cdot u_i(x) \right),
    \label{eq:fourier_layer}
\end{equation}
where $R_i$ is the Kernel function parameterized in Fourier space, and $\mathcal{F}$ and $\mathcal{F}^{-1}$ denote the Fourier transform and the inverse Fourier transform, respectively. $\mathcal{F}$ and $\mathcal{F}^{-1}$ are usually implemented using (inverse) Fast Fourier Transforms (FFT), which speed up computations significantly but require an evenly spaced discretization of the domain. 

In each Fourier layer, after the Fourier transform $\mathcal{F}$ operates on the input $u_i(x)$, we apply \textit{spectral filtering}. That is, we truncate the representation in frequency domain and retain only the lowest $k_\text{max}$ modes before passing to $R_i$. For an FNO with latent space dimension $w$, each $R_i$ is a tensor of shape $(w, w, k_\text{max})$: it applies a learnable linear transformation for each Fourier mode. Importantly, although only the lowest $k_\text{max}$ frequency modes are kept, information from higher frequency modes can still be recovered through the nonlinear activation functions applied in real (physical) space~\cite{FNO}. The architecture of a Fourier layer is illustrated in Fig.~\ref{fig:FNO}(b).

Crucially, since FNOs learn representations in Fourier space directly, their performance is relatively insensitive to the discretization of real-space~\cite{FNO}. In stark contrast to finite-dimensional operators and neural networks, which need to be retrained for different discretizations, FNOs are inherently \textit{discretization invariant}. Solutions parameterized by a single set of parameters can be transferred between different meshes. In particular, this property means that FNO can be trained on coarse grids but make predictions on finer discretizations.

\subsection{Learning Strategy\label{sec:data}}
\begin{table*}[]
    \centering
    \begin{tabular}{|c|c|c|c|c|c|}
     \hline
       Section &  Input Function & Output Function & Function Domain & Training Data & Data Domain  \\ \hline 
     \ref{sec:Heff}  &  $H(t)$   & $H_F(t)$ & $(0, T)$ & $H(t), H_F(t)$ & $(0, T)$  \\  \hline
    \ref{sec:Ht_to_obs} & $H(t)$  &  $\left< B_i(t) \right>$ & $(0, nT)$ & $H(t), \left< B_i(t) \right>$ & $(0, nT)$   \\
     \ref{sec:obs_to_obs} & $\left< B_{\text{in}}(t) \right>$ & $\left< B_{\text{out}}(t) \right>$ & $(0, nT)$ & $\left< B_i(t) \right>$ & $(0, nT)$ \\ \hline
     \ref{sec:opgrowth} & $H(t)$ & $c(t)$ & $(0, T)$ & $H(t), \left< B_i(t) \right>, \Tr(\rho B_i)$ & $(0, nT)$ \\ \hline 
     \end{tabular}
    \caption{Summary of the learning paradigms in this work and the required training data for each task. $H(t)$ denotes the set of parameters that describe a time-period Hamiltonian. $H_F(t)$ is the Floquet Hamiltonian governing stroboscopic evolution starting at time $t$.  $\left< B_i(t) \right>$ are the time-dependent expectation values for the set of one- and two-local spin operators, as defined in Eq.~\ref{eq:local_ops}, over multiple periods. $c(t)$ is a matrix encoding operator growth, whose entries are defined in Eq.~\ref{eq:cij}.}
    \label{tab:summary}
\end{table*}
In order to train Fourier Neural Operators, we need data: namely, pairs of functions of time. The data required for each learning paradigm is listed in Table~\ref{tab:summary}. In this section, we define what data is needed and briefly outline how it is obtained. For concreteness, we will be restricting to quantum systems with $L$ spin-halves (qubits). However, our approach can be naturally extended to systems with higher spins (qudits).

For each time-periodic Hamiltonian $H(t)$, we compute two quantities: \textit{effective Floquet Hamiltonian} $H_F(t)$ and expectation values of local observables, which we denote as $\left< B_i(t) \right>$. The dataset in this work consists of tuples $(H(t), H_F(t), \left<B_i(t) \right>)$ for different $H(t)$.

To obtain the effective Floquet Hamiltonians, we first numerically compute the propagators for one period, for each initial time $t \in (0, T)$:
\begin{equation}
    U(t+T, t) = \hat{\mathcal{T}} e^{-i \int_{t}^{t+T} H(t') dt'},
    \label{eq:evolution}
\end{equation}
where $\hat{\mathcal{T}}$ denotes the time-ordering operation. The effective Floquet Hamiltonian is defined as:
\begin{equation}
    H_F(t) := \frac{i}{T} \log(U(t+T, t)).
    \label{eq:HF}
\end{equation}
$H_F(t_0)$ is an effective description of the system in the sense that all measurements made at times $t = t_0 + nT$ ($n \in \mathbb{Z}$ is an integer) in two systems governed by $H(t)$ and $H_F(t_0)$ will be identical.

For efficiency of storage and evaluation, we do not store the Hamiltonians as $2^L \times 2^L$ matrices, where $L$ is the system size. Instead, $H(t)$ and $H_F(t)$ are both represented by weights in the local Pauli basis, which we now define.

The full Pauli basis for a system with $L$ qubits consists of tensor products of single-site Pauli operators:
\begin{equation}
    P := \{ \sigma_1 \otimes \sigma_2 \otimes... \otimes \sigma_L \},
\end{equation}
where each $\sigma_i \in \{ X_i, Y_i, Z_i, I_i \}$ is one of the four Pauli matrices acting on site $i$ and is different from the nonlinear activation function $\sigma$ in Eq.~\ref{eq:neural_op}. The full Pauli basis has $4^L$ elements and forms an orthonormal basis for the space of Hermitian operators acting on the Hilbert space associated with the spin chain. 

Since $P$ is exponentially large, here we restrict to operators that are at most $k$-local i.e., we only focus on Pauli strings that act non-trivially on at most $k$ consecutive sites. We denote the $k$-local basis as $P_{k}$. In contrast to $P$, $P_k$ only contains polynomially many elements, $|P_k| = \mathcal{O}(L^k)$.

The Hamiltonians $H(t)$ and $H_F(t)$ can be decomposed as linear combinations of operators in this basis. We can write:
\begin{equation}
    H(t) = \sum_j \alpha_j(t) B_j, \, B_j \in P_k,
\end{equation}
 where the coefficients $\alpha_j(t)$ are given by:
\begin{equation}
    \alpha_j(t) = \frac{1}{2^L} \Tr \left(H(t) B_j \right),
\end{equation}
and similarly for $H_F(t)$. Under this parameterization, the Hamiltonians can be represented by vectors consisting of the coefficients $\alpha_j(t)$:
\begin{equation}
    H(t), H_F(t) \longleftrightarrow \overrightarrow{H}(t), \overrightarrow{H_F}(t),
    \label{eq:H_to_vec}
\end{equation}
thereby circumventing the need to store exponentially many entries. In this work, we represent $H_F(t)$ as its weights in the three-local basis $P_{k=3}$.

Importantly, since $H_F(t)$, $H(t)$, and all $B_j \in P_k$ are Hermitian, all elements in the vector representations $\overrightarrow{H}(t)$, $\overrightarrow{H_F}(t)$ are real. This allows us to use a \textit{real} Fourier neural operator for learning.

While this parameterization is efficient, it is implicitly assuming that the effective Floquet Hamiltonians $H_F(t)$ are local. Rigorously speaking, the Floquet Hamiltonian for a nonintegrable quantum many-body system is generically highly nonlocal in the absence of many-body localization. However, in the high-frequency regime, a local prethermal Hamiltonian does exist and governs the stroboscopic evolution up to exponentially long times~\cite{Else2017}. Thus, the local $H_F(t)$ introduced above should be regarded as a prethermal Hamiltonian in the high-frequency regime.  As one decreases the driving frequency, 
truncation up to $k$-local basis with a larger $k$ may be needed to represent $H_F(t)$ faithfully. When we prepare the training dataset, we have explicitly computed the difference between the exact Floquet Hamiltonian and the truncated effective Hamiltonian and have ensured that this representation is accurate. 

The remaining data that are needed for the learning tasks are the expectation values of local observables over multiple driving periods. In this work, we focus on the set of one- and two-local spin operators, defined as:
\begin{equation}
  \{B_j\} := \{ \sigma_i^\alpha, \sigma_i^\alpha \sigma_{i+1}^{\alpha'}\},
  \label{eq:local_ops}
\end{equation}
where $i \in (1, 2, ..., L)$ is the site index, and $\alpha, \alpha' \in \{x, y, z\}$ labels the spin operator on each site. 

In this work, we will use numerically simulated expectation values to train FNOs. However, we note that the data we require, at larger system sizes, can be readily accessible from experiments performed on various platforms, such as trapped ions, ultra-cold atoms, and superconducting qubit arrays~\cite{trapped_ion1,trapped_ion2, superconducting1, superconducting2, superconducting3}. More recently, quantum computers have also emerged as a promising platform for measuring dynamics of these local observables~\cite{quantumcomputer1, quantumcomputer2}. 

To obtain the expectation values, we first prepare the system in a random product state:
\begin{equation}
    \ket{\psi(t=0)} = \bigotimes_i \ket{\psi_i},
\end{equation}
where each $\ket{\psi_i}$ is a random two-level state on site $i$ sampled from the Haar measure.

Next, we compute the expectation values of all local observables at each time:
\begin{equation}
    \left< B_i(t) \right> = \Tr \left( \rho B_i(t) \right),
\end{equation}
where $\rho = \ket{\psi(t=0)}\bra{\psi(t=0)}$. Here $B_i(t)$ is the time-evolved operator in the Heisenberg picture:
\begin{equation}
    B_i(t) = U^\dagger(t, 0) B_i U(t, 0),
\end{equation}
where $U(t, 0)$ is the propagator from time $0$ to $t$, defined in Eq.~\ref{eq:evolution}.

From the time-evolved operators $B_i(t)$, we define a matrix $c(t)$ that encodes operator growth in the system.  Each time-evolved $B_i(t)$ can be expanded in the original basis as
\begin{equation}
    B_i(t) = \sum_j c_{ij}(t) B_j,
\end{equation}
where the expansion coefficients are overlaps between $B_i(t)$ and the original basis elements,
\begin{equation}
    c_{ij}(t)  = \frac{1}{2^L}\Tr\left( B_j B_i(t) \right), B_j \in P_{k=2}.
    \label{eq:cij}
\end{equation}

Although $c(t)$ only captures how $B_i(t)$ is decomposed in the local basis, it contains valuable information about how initially simple operators develop non-local components under dynamics. In particular, the spreading of operator weight encoded in $c(t)$ can signal the onset of scrambling, thermalization, or quantum chaos.

Remarkably, $c(t)$ can be used to \textit{extrapolate} dynamics to timescales beyond the training window. Due to the composition property of unitary evolution, $c(t)$ approximately satisfies the following relation:
\begin{equation}
    c(t + NT) \simeq c(t) c(T)^N,
    \label{eq:c_extrapolation}
\end{equation}
which we prove in Appendix.~\ref{app:ctrelation}. 

Once $c(t)$ within a single period, $t \in (0, T)$, is known, the above relation allows for predicting operator growth at much longer times. By exploiting this property, we can extract physical quantities of interest, such as autocorrelation functions, from $c(t)$, at times well beyond the temporal window accessible from the data. 

In this way, short-time training data can be used to infer longer-time dynamics and observables of the quantum system. This feature makes $c(t)$ a powerful surrogate, which connects experimentally or numerically accessible, short-time measurements to long-time physical predictions, allowing us to extrapolate to timescales that would otherwise remain inaccessible due to limited quantum coherence times or prohibitive classical simulation costs.

\subsection{Physical Model \label{sec:model}}
To numerically demonstrate the validity of our approach, in this work, we focus on a concrete model: the periodically driven transverse-field Ising model (TFIM) with spatio-temporal disorder. Periodically driven TFIM is an archetypal model for exploring non-equilibrium quantum dynamics. Its various generalizations have been shown to host many novel phenomena, including dynamical freezing, entanglement transition, and time-crystalline behavior~\cite{TFIM, TFIM_2, TFIM_3, claasenflow, TFIM_harmonic_EE, TFIM_harmonic_magnetization, TFIM_harmonic_entanglementtransition}. 

The Hamiltonian of our model is given by:
\begin{equation}
    H(t) = \sum_{i=1}^L J_i X_i X_{i+1} + A_i \cos(\omega t) Z_i + h_x(i, t) X_i,
    \label{eq:TFIM}
\end{equation}
where $L$ is the number of spins/qubits, $X_i$($Z_i$) is the Pauli $x \, (z)$ operator on site $i$, $J_i$'s are the nearest-neighbor couplings, $A_i$'s are the driving strengths, and $\omega = 2 \pi /T$ is the frequency. Note the couplings and driving strengths may vary from site to site. Periodic boundary conditions are imposed, although the system is not translationally invariant. 

Due to the random parallel field $h_x(i, t)$, our model is \textit{not} integrable. In most previous studies, such integrability-breaking disorder is inhomogeneous in \textit{either} space or time~\cite{claasenflow, TFIM_disorder, TFIM_disorder2, TFIM_disorder3}. Here, however, we allow the disorder to be \textit{spatio-temporal}: the disorder may depend on both the site and time. Importantly, we do \textit{not} assume the integrability breaking disorder to be periodic in time. We can reduce the model to possess only spatial or temporal disorder by removing the dependence of $h_x$ on time and spatial index, respectively. Furthermore, the family of Hamiltonians we focus on is continuously and harmonically driven. In contrast to kicked Floquet systems, continuously driven quantum systems are rarely analytically tractable, because their Floquet propagators are fully time-ordered exponentials of $H(t)$ over one period, as opposed to products of a few exponential operators. Therefore, traditional approaches to studying disordered, continuously driven systems -- such as our model -- require expensive approximation or numerical computations.

To generate the data required for training, we first sample $H(t)$ from the family of Hamiltonians in Eq.~\ref{eq:TFIM}. The nearest-neighbor coupling $J_i$'s and the driving strengths $A_i$'s are independently and identically drawn from a uniform distribution on $[0, J]$ and $[0, A]$, respectively. The parallel fields $h_x(i, t)$ are also randomly sampled for each site $i$ and time $t$ from a uniform distribution on $[-\eta, \eta]$, where $\eta$ is the disorder strength. Each set of these coefficients $\{J_i, A_i, h_x(i, t) \}$ defines a harmonically driven, disordered TFIM. For each $H(t)$, we compute $H_F(t)$ and $\left< B_j(t) \right>$, as detailed in Sec.~\ref{sec:data}.

There is one practical note for working with the above functions (namely, $H(t), H_F(t), \left< B_i(t) \right>, c(t)$) numerically. We discretize each period $(0, T)$ into $N$ points, forming a uniform grid $ \{t_0, t_1, ..., t_{N-1} \}$, where $t_n = n \Delta t$ and $\Delta t = T/N$. The discretization is chosen to be uniform to allow for efficient implementations of the (inverse) FFT in the Fourier layers, and all quantities are evaluated only at these grid points. Importantly, while the training data are generated at the resolution of $N$ points per period, the learned FNO is not tied to it. Due to the FNO's discretization invariance, the operator can make predictions and transfer learning between different grid sizes.

To generate training data and benchmark results (which we will refer to as the ``exact'' results throughout this paper), we employ a stepwise (Trotterized) evolution method with $N=200$ points per period. Specifically, the unitary propagator, $U(t_i, t_{i+1}) = \exp(-i H(t_i) \Delta t)$, is constructed at every time $t_i$ in the discretized domain. Both Floquet Hamiltonians and local observables are then computed exactly using the unitary propagator. This approach ensures high numerical fidelity, and we have also verified convergence with respect to $N$. For data spanning $n$ periods, the number of points in the discretized domain is $nN$. Due to the significant computational cost associated with this method, our primary analysis focuses on a system size of $L=8$. However, in Appendix.~\ref{app:L14}, we present numerical results and runtime analysis for a larger system $L=14$.

In the following three sections, we will demonstrate the applications of FNOs in three learning paradigms using the above data. The hyper-parameters used for each case, including parameters of the FNO and number of training samples, are listed in Appendix~\ref{app:FNO}.

\section{Learning Effective Floquet Hamiltonians   \label{sec:Heff}}
One of the main goals of Floquet engineering is to use periodic driving as an external ``knob'' to shape quantum dynamics and design out-of-equilibrium quantum phenomena. A central object in Floquet engineering is the effective Floquet Hamiltonian $H_F$, which encapsulates the effects of driving and governs stroboscopic evolution of the system over multiple periods. Knowing $H_F$ is important for both experimental design and theoretical understanding. However, computing the Floquet Hamiltonian exactly is notoriously difficult. In the high-frequency regime, the Magnus expansion already provides an accurate and well-established approximation~\cite{magnus}. Nonetheless, exploring complementary, data-driven approaches remains valuable, particularly for systems involving complex drive protocols and spatio-temporal disorder.

As the first example of the applicability of FNO to Floquet systems, we show that FNO can learn a mapping from the time-periodic Hamiltonians to the corresponding effective descriptions, thereby learning the system's dynamics at all timescales. Furthermore, we demonstrate the unique advantages of operator learning: in particular, we show that once trained, the FNO can not only predict $H_F(t)$ for unseen Hamiltonians $H(t)$, but also \textit{generalize} beyond the driving frequency and discretization on which it is trained.

\subsection{Effective Floquet Hamiltonian and Micromotion Operator}
As discussed in Sec.~\ref{sec:background}, stroboscopic dynamics under a time-periodic Hamiltonian $H(t) = H(t+T)$ can be described by the \textit{Floquet Hamiltonian}, defined in Eqs.~\ref{eq:evolution}, \ref{eq:HF}. From its definition and the periodicity of $H(t)$, one can see that $H_F$ is also periodic in the initial time, i.e., $H_F(t_0) = H_F(t_0 + T)$ for all $t_0$. The Floquet Hamiltonians starting at different times share the same quasi-energy spectrum. In fact, they are related by unitary transformations:
\begin{equation}
    H_F(t_1) = U(t_1, t_0) H_F(t_0) U^\dagger(t_1, t_0), \label{eq:Ut0t1}
\end{equation}
where the unitaries $U(t_1, t_0)$, named the \textit{micromotion} operators, govern fast dynamics within a single period. 

Importantly, $U(t_1, t_0)$ between two times $t_1$ and $t_0$ can be computed given $H_F(t_1)$ and $H_F(t_0)$. Because the effective Hamiltonians are Hermitian, there is an orthonormal basis in which they are diagonal. Furthermore, since different $H_F$'s have the same spectrum, we can write: $H_F(t_0) = P_0 D P_0^\dagger; H_F(t_1) = P_1 D P_1^\dagger$, where $P_0$, $P_1$ are unitary, $D$ is a diagonal matrix whose entries are the quasi-energies.
Using Eq.~\ref{eq:Ut0t1}, we have $[D, P_1^\dagger U(t_1,t_0) P_0] = 0$. When $D$ has non-degenerate spectrum, $P_1^\dagger U(t_1,t_0)P_0 =J$, with
diagonal matrix $J = \diag(e^{i\theta_1}, e^{i\theta_2},\dots)$ consisting of undetermined phases. Then $U(t_1,t_0) = P_1 J P_0^\dagger$.
On the other hand, the columns of $P_{i}$ $(i=0,1)$ are eigenstates of $H_F(t_i)$, which are determined up to arbitrary phases. Thus, one can absorb the phases of $J$ into the definitions $P_0$ or $P_1$. Namely, we can choose $U(t_1,t_0) = P_1 P_0^\dagger$. When $D$ has degenerate spectrum, the unitary matrix $J$ becomes block diagonalized according to the distinct eigenvalues of $D$. One can again choose $J$ to be the identity because the eigenvectors, which are columns of $P_i$, are determined up to unitary transformations within the subspace labeled by the same quasi-energy.

Therefore, complete information about the system is encoded in $H_F(t)$, with $t \in (0, T)$~\cite{micromotion}. The ``slow'' or stroboscopic dynamics occurring at integer multiples of the period starting at any $t_0$ is described by $H_F(t_0)$. The ``fast'' dynamics within a single period is encoded in the micromotion operators. Learning how $H_F(t)$ depends on the initial time $t$ allows us to compute the propagator between two arbitrary times and learn system dynamics on all time scales.

\subsection{FNO Performance}
Traditionally, the effective Hamiltonian $H_F(t)$ is approximated using analytical techniques such as the Magnus expansion, which generates a perturbative series in the inverse driving frequency~\cite{magnus, magnus1}; a more detailed discussion on the Magnus expansion is in Appendix.~\ref{app:magnus}. In the high-frequency (prethermal) regime, low-order Floquet-Magnus truncations can already provide an accurate description of stroboscopic dynamics, and higher orders can be obtained efficiently using automated or computer-assisted implementations~\cite{magnus_automated}. Moreover, for common harmonic drives, the Magnus expansion yields an explicit $\omega$-dependent effective Hamiltonian~\cite{magnus_frequency_space}. Varying the driving frequency therefore amounts to re-evaluating the same analytic expressions rather than repeating a numerical training procedure. From this perspective, the Magnus approach offers an established, physically transparent baseline for capturing Floquet physics.

However, there are several practical considerations that motivate a complementary, data-driven approach. First, the above frequency-transferability is most direct for simple drive families with explicit analytic structure. While the Magnus expansion may be transferred across disorder realizations and parameter choices, analytical evaluation could remain challenging for more general protocols, especially those involving spatio-temporal disorder, for which one typically loses a closed-form $\omega$ dependence. Second, although micromotion operators can be obtained within the same high-frequency expansion framework~\cite{prx_micromotion}, the FNO outputs $H_F(t_0)$ for all reference times simultaneously, in a single forward pass.

Motivated by these considerations, we adopt an operator-learning viewpoint and treat finding $H_F(t)$ as an operator learning problem. $H(t)$ and $H_F(t)$ can both be viewed as functions acting on the domain, namely mappings from $ D = (0, T)$ to $\mathbb{H}_{d_H}$, where $\mathbb{H}_{d_H}$ denotes the space of all $d_H \times d_H$ Hermitian matrices. In this language, the Magnus expansion (after truncation) provides an analytic approximation to $H_F(t_0)$, whereas learning $H_F(t)$ 
aims to infer a single time-resolved operator-valued function over the entire period. This enables predicting the effective Floquet Hamiltonian for any reference time within the drive cycle.

In this formalism, the FNO aims to learn an operator mapping from $H(t)$ to $H_F(t)$:
\begin{equation}
    \mathcal{N}: H(t) \rightarrow H_F(t), \, t \in D.
    \label{eq:NNHF}
\end{equation}
This (highly non-linear) operator is implicitly given in Eqs.~\ref{eq:evolution} and \ref{eq:HF}.

The loss function corresponding to each data point $(\overrightarrow{H}(t), \overrightarrow{H_F}(t))$ is the Mean Square Error (MSE) between the network's output and the ground truth, averaged over all times:
\begin{equation}
   \mathcal{L} = \frac{1}{N_{D}} \sum_{t_j \in D_j} \left\lVert \mathcal{N} \left(\overrightarrow{H}(t_j) \right) - \overrightarrow{H_F}(t_j) \right\rVert^2,
   \label{eq:HF_loss}
\end{equation}
where $N_D$ is the number of points in the discretized domain $D_j$. The FNO aims to minimize the sum of the above loss function over all data points in the training set; see Eq.~\ref{eq:NO_loss}.

\begin{figure}
    \centering
    \includegraphics[width=1\linewidth]{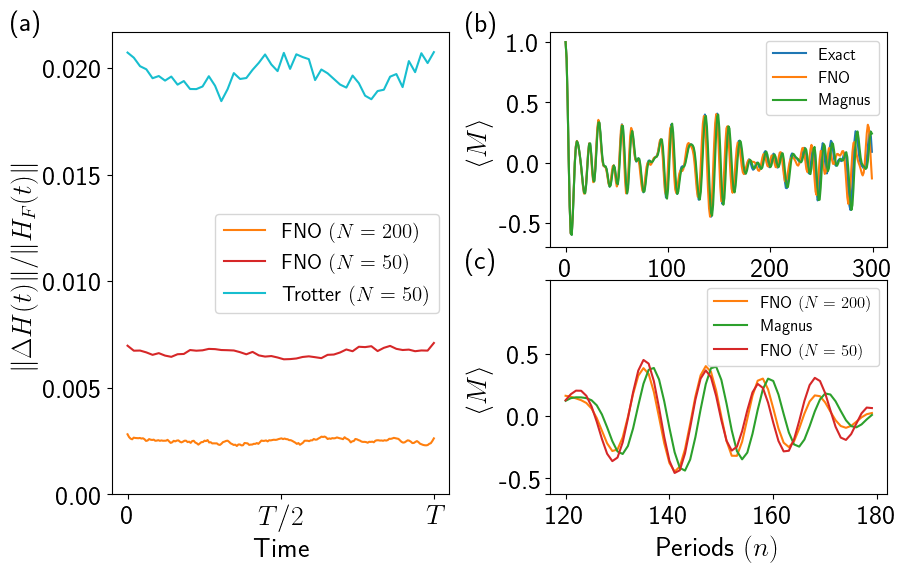}
    \caption{Demonstrations of FNO's discretization invariance and high accuracy in Floquet Hamiltonian learning. \textbf{(a)} Frobenius norms of the differences $\Delta H(t)$ between the exact Floquet Hamiltonian $H_F(t)$ and three approximations: Trotterization using $N=50$ points; predicted effective Hamiltonian $\mathcal{N}(H(t))$ by an FNO trained on $N=200$ points; and prediction by an FNO trained on $N = 50$ points. Each difference is normalized by $ \lVert H_F(t) \rVert$. \textbf{(b)} Stroboscopic measurements of magnetization (Eq.~\ref{eq:magnetization}) under the exact Floquet Hamiltonian, FNO-predicted effective Hamiltonian, and the Magnus expansion. \textbf{(c)} Intermediate-time stroboscopic evolution of the magnetization under Floquet Hamiltonians given by the Magnus expansion, by an FNO trained on $N=50$ points, and an FNO trained on $N=200$ points. All panels use the same parameters: $L = 8, A/J = 0.5,$ $\eta/J = 0.1, \omega/J = 20$.}
    \label{fig:Heff_performance}
\end{figure}

We first test the generalization ability of our neural operator by considering a Hamiltonian $H(t)$ of the form given by Eq.~\ref{eq:TFIM}, but with coefficients that have \textit{not} been seen during training. We benchmark the performance of the FNO by examining differences between the actual Floquet Hamiltonian $H_F(t)$ and Floquet Hamiltonians obtained from three methods: predicted $H_F(t)$ by FNO trained on the full grid $(N=200)$ (orange); predicted $H_F(t)$ by FNO trained and evaluated on a coarser grid $(N=50)$ (red); Trotterized approximation using $N=50$ points (cyan). In Fig.~\ref{fig:Heff_performance}(a), we show the Frobenius norm of $\Delta H(t)$ for the three cases. 
The relative error from the FNO's predictions, even trained on a coarse grid, consistently remains within $0.8 \%$ over the period, indicating accurate learning of the Floquet Hamiltonian. Furthermore, predictions by the FNO trained on $N=50$ points outperform Trotterization on the same mesh, due to the discretization invariance of operator learning.

We also demonstrate the validity of the FNO's learning by computing stroboscopic expectation values of observables. In particular, we focus on the system's magnetization, which is defined as
\begin{equation}
    M = \frac1L \sum_{i=1}^L Z_i.
    \label{eq:magnetization}
\end{equation}
The system is initially prepared in a product state in which all spins are in the $\ket{\up \,}$ eigenstate of the $Z$ operator, $Z_i \ket{\up \,} = \ket{\up \,}$. 

In Fig.~\ref{fig:Heff_performance}(b), we show the stroboscopic evolution of magnetization under the actual Floquet Hamiltonian $H_F(t)$, the effective Hamiltonian learned by the neural operator $\mathcal{N}(H(t))$, and the effective Hamiltonian given by the second-order Magnus expansion, all on the same discretization of $(0, T)$. For a brief review of the Floquet-Magnus expansion, see Appendix.~\ref{app:magnus}. The Floquet Hamiltonian learned by FNO is capable of capturing stroboscopic dynamics accurately for more than $200$ periods. At short and intermediate times, FNO has performance comparable to the Magnus expansion. However, at later times, the Magnus expansion approximates the dynamics better than the FNO. The reason is that second-order Magnus expansion yields a Hamiltonian with a larger spatial support (containing terms up to $4$-local) than the one learned by the FNO, which is at most $3$-local due to our parameterization.

With the validity of our approach confirmed, we next explore the unique advantages of operator learning. As discussed in Sec.~\ref{sec:fno}, because FNOs are parameterized in Fourier space directly, the learning performance is minimally dependent on how the real space (in this case, time) is discretized. This so-called property of discretization invariance or zero-shot super-resolution means that the neural operator can be trained on a coarse grid and evaluated on a finer one~\cite{FNO}. As shown in Fig.~\ref{fig:Heff_performance}(a), FNO on a coarse grid outperforms equal-mesh Trotterization in terms of relative error. In Fig.~\ref{fig:Heff_performance}(c), we further demonstrate this property by showing stroboscopic evolution of $M$ under three Hamiltonians: the Hamiltonian from Magnus expansion, the effective Hamiltonian predicted by an FNO trained on the full grid ($N=200$ points), and one outputted by an FNO trained on a grid four times coarser ($N=50$ points). Although the predicted $H_F$ from the coarser grid is less accurate than $H_F$ from the finer discretization, it nonetheless captures the dynamics reasonably well, up to hundreds of periods.

\begin{figure}
    \centering
\includegraphics[width=\linewidth]{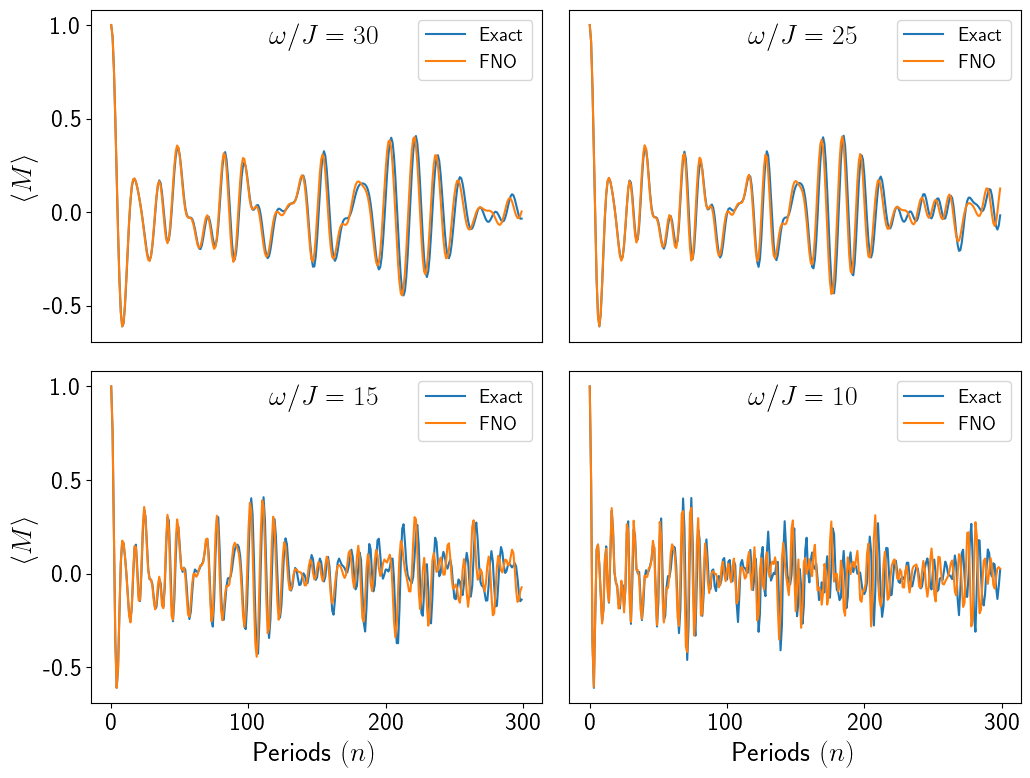}
    \caption{Transferability of FNO across driving frequencies, \textit{without retraining}. The FNO, trained on data from a system with driving frequency $\omega/J = 20$, is used to predicted Floquet Hamiltonians for systems at other frequencies $(\omega/J = 35, 25, 15, 10)$. For each extrapolated frequency, the stroboscopic time evolution of magnetization under the exact Floquet Hamiltonian and the FNO's prediction is shown. The FNO's predictions remain valid in the high-frequency regime. Other parameters used for the plots are: $L=8, A/J = 0.5, \eta/J = 0.1$.  }
\label{fig:frequency_extrapolation}
\end{figure}

Besides being invariant under discretization of the domain, neural operator are also naturally agnostic to the domain itself. This is because operator learning reconstructs continuous mappings between function spaces and does not depend on the underlying domain. It has been shown that neural operators can adapt to and transfer learning between different boundary conditions, geometries, and domains~\cite{geometry_transfer1, geometry_transfer2, geometry_transfer3, geometry_transfer4}. 

Here, we exploit the domain-agnostic nature of FNO by testing its ability to generalize beyond the driving frequency $\omega$ used in training. Changing the frequency from $\omega$ to $\omega'$ amounts to a rescaling of the domain, $D = (0, 2\pi / \omega) \rightarrow D' = (0, 2\pi/\omega')$. As shown in Fig.~\ref{fig:frequency_extrapolation}, the FNO is able to predict effective Hamiltonians for frequencies outside the training set. As long as the driving frequency is still large compared to other energy scales, so that $H_F$ remains well-represented by the local basis, the FNO can extrapolate to a broad range of frequencies, \textit{without retraining}. In the high-frequency limit, the predicted Floquet Hamiltonians govern stroboscopic dynamics accurately for hundreds of cycles. At $\omega \lesssim 10J$, the predictions begin to become inaccurate.

The two examples above clearly demonstrate the discretization invariant and domain agnostic nature of operator learning and highlight its unique advantage. Traditional numerical approaches, such as computing $H_F$ by brute force or solving flow equations~\cite{flowequation1, flowequation2}, are naturally tied to an underlying grid or discretization of the domain. Furthermore, those methods require re-computations when $H(t)$ and the driving frequency change. In contrast, once the FNO is trained, it is able to evaluate the effective Hamiltonian at all starting times, for a wide range of Hamiltonian parameters, frequencies, and underlying meshes, via a single forward pass through the network, with accuracy comparable to these traditional methods but with lower computational overhead.

\section{Learning Expectation Values \label{sec:observable}}
In quantum many-body systems, computing expectation values is generally a challenging task and requires the knowledge of full, exponentially large wavefunctions. When time evolution over a long window is involved, it becomes even more memory and resource intensive to numerically keep track of the wavefunctions at each time. In fact, even when full wavefunctions or efficient representations thereof (such as the matrix-product states) are available, it is still often costly to compute expectation values.

In this section, we show that FNO is able to circumvent those numerical hurdles and predict expectation values of local observables over multiple cycles directly, with remarkable accuracy. While the local observables do not encode all information about a quantum system, they are nonetheless practically and experimentally relevant quantities. 

Remarkably, we demonstrate that FNOs can be trained to predict expectation values in two complementary ways: using either Hamiltonian parameters $H(t)$ as inputs, or from \textit{partial} observations of the system, i.e., using only a \textit{subset} of the local observables as inputs. The two ways of learning are applicable to different situations, making the FNO a versatile tool suitable for studying quantum dynamics from a wide variety of experimentally accessible data.

\subsection{Expectation Values from Hamiltonian Parameters \label{sec:Ht_to_obs}}
We first show that FNOs are able to learn mappings from Hamiltonian parameters to expectation values of local observables i.e., they can learn the following operator:
\begin{equation}
    \mathcal{N}: H(t) \rightarrow \{ \left< B_i(t) \right> \}.
    \label{eq:Ht_to_observables}
\end{equation}

The FNO is trained on tuples of data $\left( \overrightarrow{H}(t), \overrightarrow{\left<  B_i(t)\right>} \right)$. Here $\overrightarrow{H}(t)$ is the vector representation of $H(t)$ (see Eq.~\ref{eq:H_to_vec}), and $\overrightarrow{\left<  B_i(t)\right>}$ is a vector consisting of expectation values of local observables at time $t$. All expectation values are taken with respect to the same initial random product state $\ket{\psi_0}$, constructed by independently sampling a two-level state from the Haar measure on each site.

The loss function corresponding to a single data point is again the Mean Squared Error (MSE) between the FNO's prediction and the ground truth, averaged over all times in the discretized domain $D_j$:
\begin{equation}
      \mathcal{L} = \frac{1}{N_D} \sum_{t_j \in D_j} \left\lVert \mathcal{N}\left(\overrightarrow{H}(t_j) \right) - \overrightarrow{\left<  B_i(t_j)\right>} \right\rVert^2,
\end{equation}
and the FNO aims to minimize the above loss function over all samples in the training set.

\begin{figure}
    \centering
     \includegraphics[width=\linewidth]{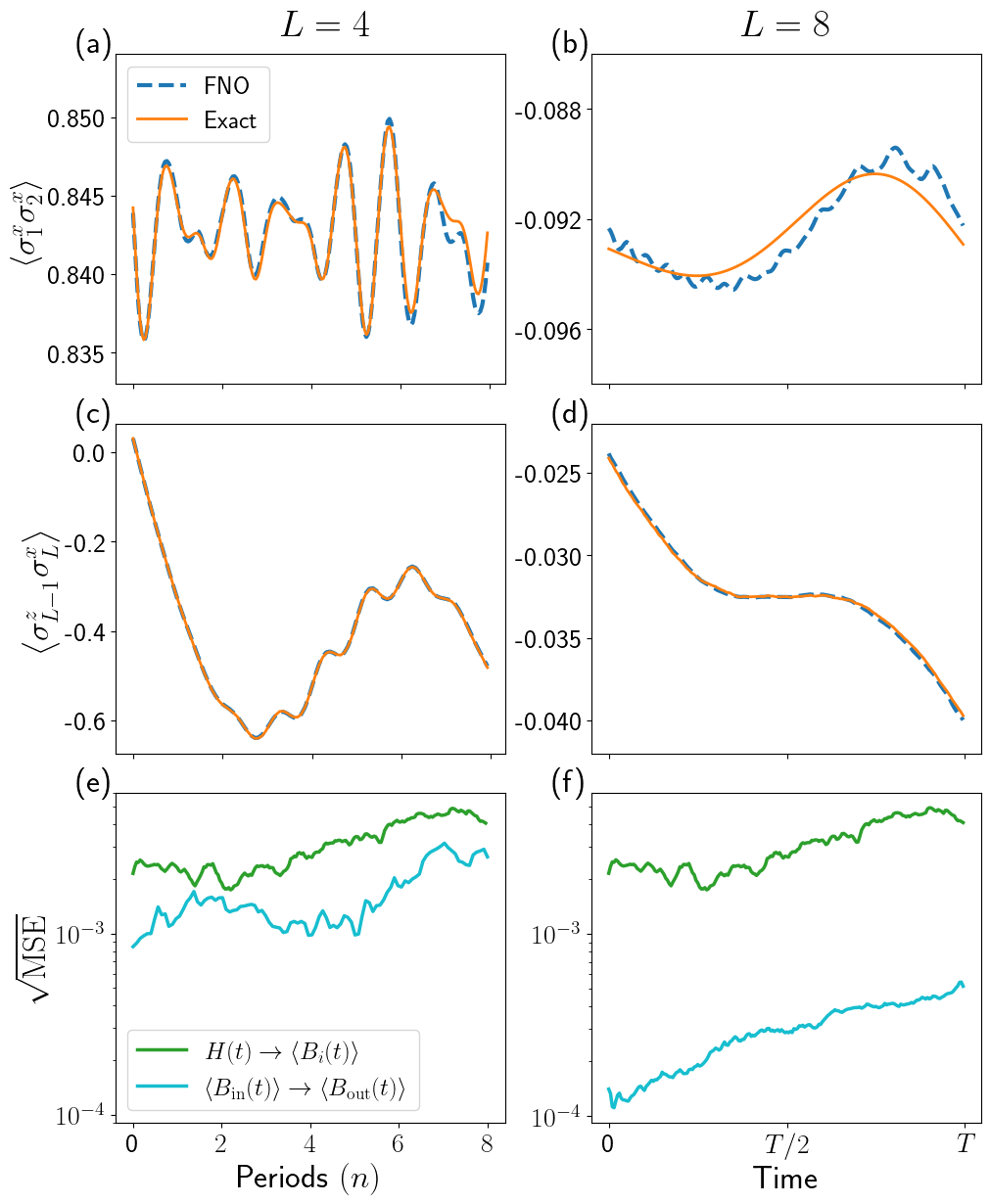}
    \caption{Performance of FNO for learning expectation values of local observables. \textbf{(a, b)} FNO trained with Hamiltonian parameters as inputs, compared to the exact dynamics obtained from converged Trotterization. \textbf{(c, d)} Predicted and exact expectation values of two-local observables, using one-local observables as FNO's inputs. \textbf{(e, f)} Root Mean Squared Error (RMSE) between the FNO's predictions and ground truths at each time, averaged over all predicted observables, for the two training approaches. Using both methods, FNO achieves accurate learning of local observables, both for a single cycle ($L=8$) and over multiple periods ($L=4$). Parameters:  $A/J = 0.5, \eta/J = 0.1, \omega/J = 20$.}
    \label{fig:observable_performance}
\end{figure}

Once the FNO is trained, an unseen instance of $H(t)$ can be directly inputted to the neural operator.
Fig.~\ref{fig:observable_performance}(a, b) shows the agreement between the FNO's predictions and the exact dynamics of a local observable under an unseen Hamiltonian. The Root Mean Square Error (RMSE) averaged over all $m$ observables (indexed by $i$), defined as
\begin{equation}
    \text{RMSE}(t) := \sqrt{\frac{1}{m} \sum_{i=1}^m \lVert \mathcal{N}(H(t))_i - \left< B_i(t) \right> \rVert ^2},
    \label{eq:rmse}
\end{equation}
is shown in the green curves in Fig.~\ref{fig:observable_performance}(e, f), demonstrating accurate learning both for a single cycle ($L=8$) and over multiple periods ($L=4$).

The FNO's ability to predict observables from Hamiltonian parameters is particularly useful when one wants to explore the parameter space of a model. Instead of re-measuring or re-computing expectation values for each new set of parameters, one can instead obtain those quantities efficiently, from a single forward pass through the FNO. Our approach could also be applied to reverse-engineer Hamiltonians, i.e., to find optimal parameters of $H(t)$ that yield desired behaviors of local observables.

While previous works have explored how neural networks can predict observable dynamics~\cite{DeepLearningCircuit, random_driving, dual_complexity}, we emphasize that our approach is unique in two ways. Firstly, Fourier Neural Operators are empirically and theoretically known to be more sample efficient~\cite{boulleReview, FNO, FNO_complexity} than other deep learning approaches. That is, to achieve the same accuracy, operator learning approaches generally require fewer samples. In this work, the training data set required to learn the mapping in Eq.~\ref{eq:Ht_to_observables} is orders of magnitude smaller than those in previous works; see Table.~\ref{tab:hyperparameters}. Generating the data necessary for training FNOs, through either simulations or experiments, is therefore much less demanding in both time and hardware (such as disk memory). Secondly, our physical model (Eq.~\ref{eq:TFIM}) is highly disordered and non-integrable, but the FNO still manages to learn with high accuracy. This is again due to the unique properties of operator learning, which has been shown to be highly robust under noise~\cite{Kovachki_NeuOp, FNO, NO_noise, NO_noise2}. Among the various neural operator architectures, the FNO is particularly suitable for treating disordered systems, because spectral filtering (truncation in the frequency domain) in FNOs further removes the high-frequency noise from the signals~\cite{FNO, NO_noise2}. Because of these advantages, the FNO may be more suitable than traditional neural networks as a tool to study more realistic physical models with driving, disorders, and impurities.

\subsection{Mapping Between Expectation Values \label{sec:obs_to_obs}}
In realistic settings, complete knowledge of the Hamiltonians is oftentimes neither possible nor necessary. Instead, quantities that are experimentally accessible and of practical interest are usually expectation values of local observables. 

In this section, we explore a novel application of deep learning techniques to quantum many-body systems: we demonstrate that FNO can learn mappings \textit{between} expectation values. Once trained, the FNO can predict the dynamics of \textit{all} local observables, with only a \textit{subset} of them as the inputs. Our approach circumvents the need to reconstruct or learn Hamiltonians altogether and instead establishes a \textit{direct} connection between experimentally accessible observables.

Consider the set of all one- and two-local observables $\{  B_i \}$, defined in Eq.~\ref{eq:local_ops}. We partition them into two disjoint sets of operators, labeled as $\{ B_{\text{in}} \}$ and $\{ B_{\text{out}} \}$, so that $\{ B_{\text{in}} \} \bigcup  \{ B_{\text{out}} \} = \{B_i\}$. The FNO is trained to learn the mapping between these sets of expectation values:
\begin{equation}
    \mathcal{N}: \{\left< B_{\text{in}}(t) \right> \} \rightarrow \{\left< B_{\text{out}}(t) \right> \}
\label{eq:observables_to_observables}
\end{equation}
by minimizing the Mean Squared Error (MSE) between the FNO's predictions and the actual expectation values for the output observables
\begin{equation}
 \mathcal{L} = \frac{1}{N_D} \sum_{t_j \in D_j} \left\lVert \mathcal{N}\left(\overrightarrow{\left<  B_{\text{in}}(t_j)\right>} \right) - \overrightarrow{\left<  B_{\text{out}}(t_j)\right>} \right\rVert^2
 \label{eq:observable_to_observable_loss}
\end{equation}
over all data points in the training set. Here $N_D$ again denotes the number of points in the set $D_j$. Similar to Sec.~\ref{sec:Ht_to_obs}, $\overrightarrow{\left<  B_{\text{in/out}}(t_j) \right>}$ denotes the vector whose entries are expectation values of operators in $B_\text{in/out}$ at time $t_j$.

There is a subtle but important caveat with using expectation values as inputs to the FNO. Unlike the two previous applications (Eqs.~\ref{eq:NNHF},~\ref{eq:Ht_to_observables}), the input functions here are not strictly periodic in time. In Appendix.~\ref{app:padding}, we explore different ways to remedy the associated edge effects. Throughout this section, we adopt zero-padding at the end of each input sequence to mitigate the numerical artifacts.

As a concrete example of how mappings between observables work, we first demonstrate a special case of Eq.~\ref{eq:observables_to_observables}. We show that FNO can accurately learn expectation values of all two-local observables, using one-local observables as inputs. That is, we choose $\{  B_{\text{in}}\} = \{  \sigma_i^\alpha \}; \{  B_{\text{out}}\} = \{  \sigma_i^\alpha \sigma_{i+1}^{\alpha'} \}$, where $i$ labels the site and $\alpha, \alpha' \in \{x, y,z\}$. This example is also of practical interest, because measuring one-local observables is generally easier than obtaining two-local expectation values, which require measuring the correlation between neighboring sites.

As shown in Fig.~\ref{fig:observable_performance}(c, d), the FNO accurately predicts the dynamics of two-local observables, both over a single cycle and for multiple periods. Importantly, the two-body term shown in the figure is not in the family of Hamiltonians in Eq.~\ref{eq:TFIM}, highlighting that the FNO captures correlators beyond those directly encoded in the Hamiltonian. All two-local spin correlators are learned with high precision, as shown in the cyan curves in Fig.~\ref{fig:observable_performance}(e, f).

Remarkably, FNOs exhibit broader predictive power beyond this setup. The input observables do \textit{not} have to be all one-local. In general, the inputs $\{ B_{\text{in}}\}$ can be \textit{any} randomly chosen subset of $\{B_i\}$; the FNO can learn the mapping from $\{ B_{\text{in}}\}$ to the rest of the local observables, $\{ B_{\text{out}}\} = \{ B_i \} \backslash \{ B_{\text{in}}\}$. We have also empirically discovered that the number of input observables that is required for FNO to learn the dynamics is much smaller than $3L$, the number of one-local observables. In fact, we will argue that for our model, the FNO only needs $\sim L$ input observables to learn an accurate representation of the system's dynamics.

\begin{figure}
    \centering
    \includegraphics[width=\linewidth]{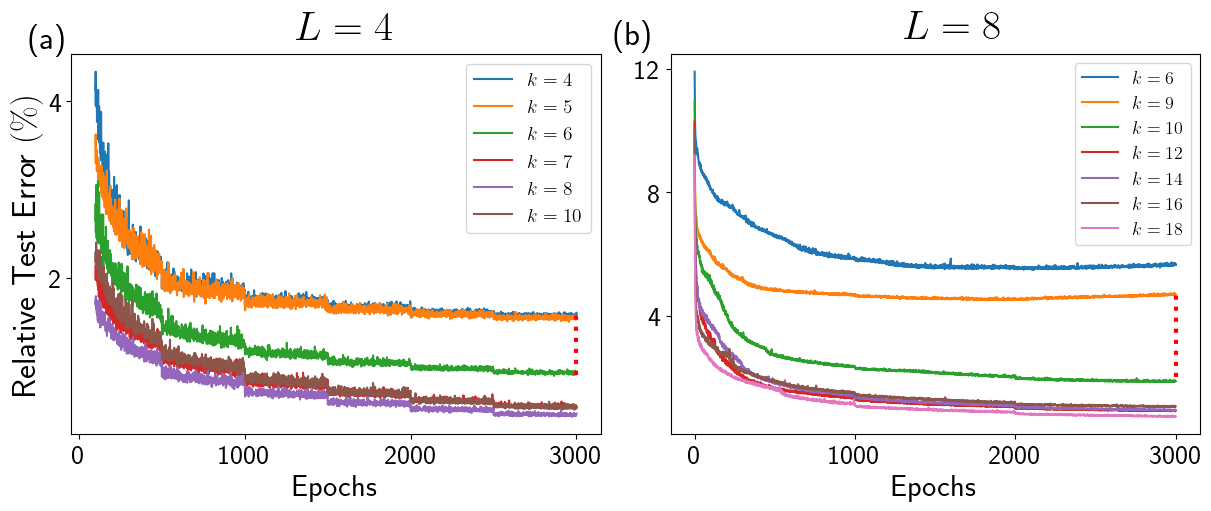}
    \caption{Relative test error during training for different numbers of input observables $k$: (a) $L=4$ (8 periods) and (b) $L=8$ (1 period). Each loss curve is averaged over $8$ independent runs. For both system sizes, a sharp drop in error occurs once $k$ exceeds $k^* = L+1$, as indicated by dashed red lines. Parameters:  $A/J = 0.5, \eta/J = 0.1, \omega/J = 20$.}
    \label{fig:learnability_transition}
\end{figure}

To demonstrate this point, we first examine the testing loss during training for different numbers of input observables $k$, where $k:= |\{ B_{ \text{in}}\}|$ is the number of observables fed into the FNO. The relative test error is defined analogously to Eq.~\ref{eq:observable_to_observable_loss}, except normalized by  the ground truth:
\begin{equation}
    \mathcal{L}_{\text{rel}}:= \frac{1}{N_D} \sum_{t_j \in D_j} \frac{\left\lVert \mathcal{N}\left(\overrightarrow{\left<  B_{\text{in}}(t_j)\right>} \right) - \overrightarrow{\left<  B_{\text{out}}(t_j)\right>} \right\rVert^2}{\left\lVert \overrightarrow{\left<  B_{\text{out}}(t_j)\right>} \right\rVert^2},
\end{equation}
and averaged over data points in the testing set (which are \textit{not} seen during training). For each $k$, we \textit{randomly} select $k$ observables from $\{B_i\}$ as the inputs and train the FNO to predict dynamics of the remaining ones. To ensure that the loss curves are not artifacts of a particular sampling from $\{ B_i \}$, we repeat training $8$ times for each $k$ and average over the loss curves. As shown in Fig.~\ref{fig:learnability_transition}, the testing loss drops sharply once $k$ exceeds $ k^* = L+1$, both for $L=4$ over multiple periods and $L=8$ over a single period.

We hypothesize that this behavior suggests a \textit{learnability transition}. We give a heuristic argument for why the transition occurs at $k^* \sim L+1$ from parameter counting. We denote the number of points in time as $N$, and the system size as $L$. Each Hamiltonian in the family (Eq.~\ref{eq:TFIM}) is specified by $NL + 2L$ \textit{learnable} degrees of freedom:  $L$ independent local couplings $J_i$, $L$ driving strengths $A_i$, and $NL$ coefficients of the spatio-temporal parallel field $h_x(i, t)$. On the other hand, with $k$ input observables, the neural operator receives $kN$ \textit{input} degrees of freedom. In order for FNO to construct an internal representation of the system, the number of input parameters should at least match the number of learnable parameters. Equating the two yields the estimate $k^* \gtrsim L+1$, consistent with our numerical findings. This indicates that FNOs are not only sample efficient, but also parameter efficient: they require relatively few inputs to capture the underlying physics.

Our results could have direct implications for applications of FNOs to experimental data from quantum computers, where the system sizes exceed the reach of classical numerical methods such as exact diagonalization. If the observed scaling persists at larger system sizes $L$, an FNO could be used to make accurate predictions on the quantum system's local dynamics using measurements of only a subset of observables, whose number scales \textit{linearly} with the system size. Furthermore, from the above argument, for Hamiltonian of lower complexity than our model (fewer learnable parameters), the number of required input observables might be even smaller.

\section{Learning Operator Growth \label{sec:opgrowth}}
As the last example of how FNOs can be applied to understanding quantum dynamics, we go beyond expectation values and study operator growth in time-periodically driven quantum systems. Understanding how initially local operators evolve and spread in time is key to probing the underlying structure of the quantum system: for example, whether they are integrable, chaotic, or localized. Characterizing operator spreading also provides insights into the mechanism of thermalization, as well as information transport.

How operators grow under dynamics governed by a Hamiltonian $H(t)$ in the family described by Eq.~\ref{eq:TFIM} is not \textit{a priori} obvious. Indeed, the interplay of spatiotemporal disorder and periodic driving leads to a competition between localization and ergodicity, and the mechanism of thermalization depends on the particular realization of the parallel field $h_x(i, t)$. Therefore, in order to understand how information spreads in time, one has to obtain the matrix that encodes operator $c(t)$, defined in Eq.~\ref{eq:cij}, for each $H(t)$. Computing $c(t)$ exactly, however, is prohibitively expensive, especially for systems with more than a handful of qubits: one has to time-evolve each initial operator $B_i$, project back onto the original basis, and analyze how the weights evolve in time.

Alternatively, $c(t)$ can be obtained through experimental measurements. For a given initial state $\rho$, we can measure the time-dependent expectation values of all local operators $\langle B_i(t)\rangle$. They are related to $c(t)$ via:
\begin{equation}
    \langle B_i(t) \rangle = \sum_j c_{ij}(t) \Tr(\rho B_j).
    \label{eq:Bandct}
\end{equation}

Let us denote the number of local operators as $m$, where $m = |\{B_i\}|$, then $c$ at each time $t$ is an $m\times m$ matrix. In order to calculate $c_{ij}(t)$ from these measurements, we have to take at least $m$ different initial states $\rho^{(1)},\cdots, \rho^{(m)}$, so that we can construct an invertible matrix $\rho B$, with entries $(\rho B)_{jn} =\Tr(\rho^{(n)}B_j)$. We can then invert the above relation and obtain
\begin{equation}
    c(t) = B(t) (\rho B)^{-1},
\end{equation}
where the $i$th column of the matrix $B(t)$ consists of measurement values $\langle B_i^{(n)}(t) \rangle$ with respect to each initial state $\rho^{(n)}$.

Both of these methods are intensive in resources and computational time. More importantly, the methods only yield $c(t)$ for a particular Hamiltonian $H(t)$ and have to be repeated for another realization of the integrability-breaking field $h_x(i, t)$. In contrast, using the FNO as a surrogate model allows for extrapolation to unseen Hamiltonians and circumvents the need of re-computing or re-measuring for every disorder realization. 

To this end, in this section, we use FNO to learn the mapping from $H(t)$ to $c(t)$:
\begin{equation}
    \mathcal{N}: H(t) \rightarrow c(t).
\end{equation}

We highlight that in contrast to previous sections, the FNO is \textit{not} directly trained on tuples of $(H(t), c(t))$. Instead, the data used for training are experimentally accessible quantities: namely, expectation values of local observables $\left< B_i(t) \right>$ and partial information about the density matrix $\Tr(\rho B_i)$. We note that for each Hamiltonian $H(t)$, expectation values with respect to multiple density matrices are needed in the data set to learn $c(t)$. Indeed, if $\left< B_i(t)\right>$ is obtained for just a single initial state, $c(t)$ cannot be constrained during learning. In practice, for each Hamiltonian $H(t)$, we use $N_\rho = 20$ initial states to restrict the structure of $c(t)$. Each $\rho$ is created from a randomly drawn initial product state, $\rho = \ket{\psi}\bra{\psi}$.

As shown in Eq.~\ref{eq:c_extrapolation}, $c(t)$ satisfies an approximate composition property that relates $c(t + nT)$ to $c(t) c(T)^n$. This allows us to perform 
training using data spanning \textit{multiple} periods, while restricting the domain of input and output functions to $(0, T)$, i.e., learning $c(t)$ within a single period. To learn $c(t)$, we use the following loss function, which measures the discrepancy (MSE) between the actual observables (over multiple periods) and predicted expectation values from $c(t)$:
\begin{equation}
 \mathcal{L} = \frac{1}{n_\text{max} N} \sum_{t,n} \left\lVert \left< B_i(t+nT)\right> - \sum_{j} \left[\tilde{c}(t)\tilde{c}(T)^n\right]_{ij} \Tr(\rho B_j) \right\rVert^2,
 \label{eq:opgrowth_loss}
\end{equation}
averaged over all data points. Here $\tilde{c}(t) = \mathcal{N}(H(t))$ is the $c(t)$ matrix predicted by the FNO. The loss function is minimized over all $t \in (0, T)$ and $n \in (0, 1, \dots n_\text{max})$, where $n_\text{max}$ is the total number of periods used in the data. By minimizing the above loss function, the FNO learns the $c(t)$ matrix that best matches the experimentally measured expectation values.

\begin{figure*}
    \centering
    \includegraphics[width=1\linewidth]{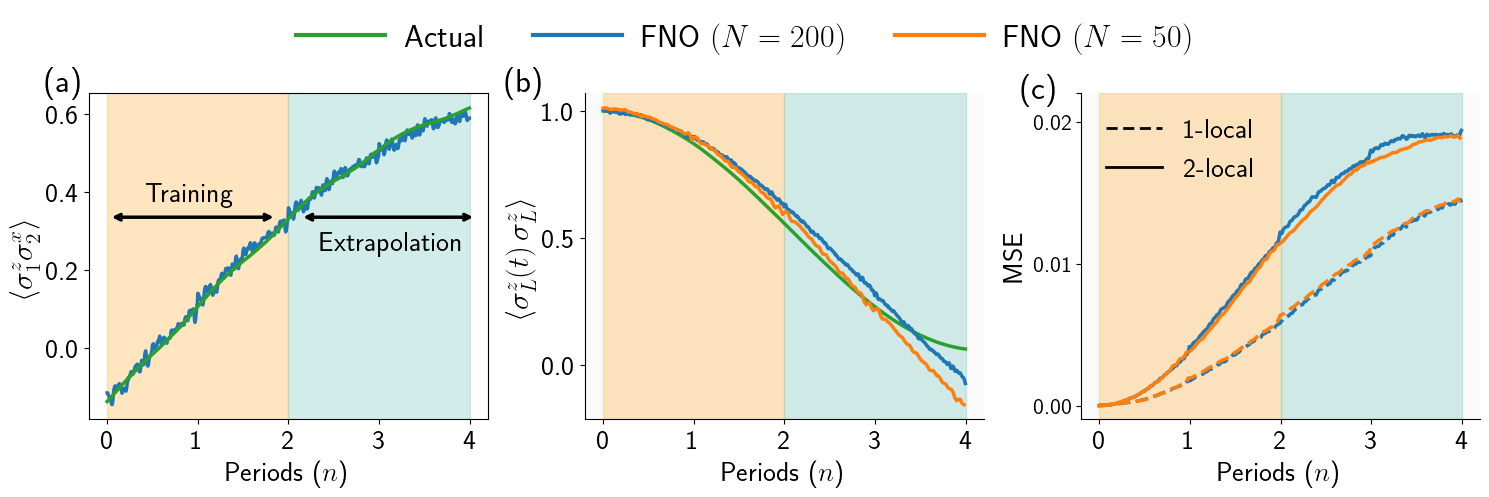}
    \caption{Performance of FNO in learning operator growth and extrapolation beyond the training window. \textbf{(a)} Predictions of expectation values of local observables under an unseen Hamiltonian $H(t)$ and an \textit{unseen} density matrix $\rho$, obtained from contracting $\tilde{c}(t)$ and partial traces $\Tr(\rho B_j)$, via Eq.~\ref{eq:Bandct}. \textbf{(b)} Exact and predicted autocorrelation function $\left< \sigma_L^z(t) \sigma_L^z \right>$ under dynamics generated by an unseen Hamiltonian. FNOs trained on both grid sizes ($N=200, 50$) reproduce the correlation accurately. \textbf{(c)} Mean-squared error (Eq.~\ref{eq:Bit_error}) in predicting operator growth, averaged over one-local (dashed lines) and two-local (solid lines) initial operators. Parameters: $n_\text{max}=2,\omega/J = 20$, $A/J=0.5, h_x/J = 0.1, L=4$.}
    \label{fig:opgrowth}
\end{figure*}

To assess the performance of $c(t)$ learned by the FNO, we first check if it is able to predict expectation values of local observables using Eq.~\ref{eq:Bandct}, for an unseen Hamiltonian. In Fig.~\ref{fig:opgrowth}(a), we show that the reconstructed expectation values, obtained by contracting the learned $\tilde{c}(t)$ with partial traces $\Tr(\rho B_j)$, closely follow the exact trajectory. This serves as the first confirmation that $\tilde{c}(t)$ learns the relevant operator dynamics, instead of memorizing time series.

Remarkably, our findings imply that FNOs can not only extrapolate to unseen Hamiltonians, but also predict expectation values for unseen \textit{initial states}. In most previous works, time-forecasting tasks have been restricted to measurements with respect to the same initial state~\cite{DeepLearningCircuit, dual_complexity, FNO}. In contrast, our approach of learning $c(t)$ requires only \textit{partial} information about $\rho$ (namely, its partial traces $\Tr(\rho B_j)$) to predict expectation values during the training window and beyond, for any random initial state. Our approach is therefore broadly applicable across various initial conditions, thereby significantly extending the capacity of machine-learning-based time-forecasting methods.

Next, we study the autocorrelation function at infinite temperature, which is defined as
\begin{equation}
    A_i(t) = \left< B_i(t) B_i(0) \right>_\infty = \Tr \left(B_i(t) B_i(0) \right).
\end{equation}

The boundary-operator version of this autocorrelation function was proposed as a tool~\cite{Yates2019,Schmid2024,Schmid2025} to detect strong modes in a spin chain. Moreover, this quantity could also measured on Google's quantum computer~\cite{Mi2022} via averaging over the autocorrelation functions from different random states, demonstrating the existence of the strong edge modes. In this work, we choose $A_i(t)$ to test the performance of FNO because $A_i(t)$ is not directly related to any expectation values in the training data, which means that the FNO cannot predict it by memorizing the dataset. Instead, it has to learn a representation of the underlying physics to produce accurate autocorrelation functions $A_i(t)$.

Since $B_i(t)$ can be decomposed in the original basis via Eq.~\ref{eq:cij}, $A_i(t)$ can be written as:
\begin{equation}
    A_i(t) = \sum_j c_{ij}(t) \Tr \left( B_j(0) B_i(0) \right) = c_{ii}(t).
\end{equation}
In the last line, we used the orthogonality of the Pauli strings under the Hilbert-Schmidt inner product, $\Tr \left( B_j^\dagger B_i \right) = \delta_{ij}$. The infinite-temperature autocorrelation function $A_i(t)$ is directly given by the diagonal entries of $c(t)$.

As shown in Fig.~\ref{fig:opgrowth}(b), the trained FNO predicts the autocorrelation function for an edge operator $\left< \sigma_L^z(t) \sigma_L^z \right>_\infty$ accurately, also for a time window beyond the training interval. Remarkably, the FNO trained on a grid four times coarser also makes accurate predictions on the autocorrelator, thanks to the discretization invariant property of FNOs. At time horizon about two times the training window, the predictions by the FNO start to deviate from the true autocorrelators.

As the final check on the learned matrix $\tilde{c}(t)$, we evaluate how well it captures operator growth. To this end, we examine the mean squared error (MSE) between the $i$th column of $\tilde{c}$ and the exact $c(t)$:
\begin{equation}
    \|\delta B_{i}(t)\| \equiv \sum_j |\tilde{c}_{ij}(t) - c^{\text{exact}}_{ij}(t)|^2,
    \label{eq:Bit_error}
\end{equation}
which quantifies the error $\tilde{c}(t)$ makes when decomposing $B_i(t)$ into the original basis.

In Fig.~\ref{fig:opgrowth}(c), we show this error, averaged over initially one-local and two-local operators. Remarkably, an FNO trained on the coarser grid still captures operator growth accurately, even beyond the training window.

We emphasize again that the training procedure relies only on short-time, experimentally measurable data and makes no assumptions about the form of the underlying Hamiltonians. These properties make the FNO architecture broadly applicable to processing data from quantum simulator platforms. Importantly, our method requires data only from a short time interval, making it particularly practical, as current experimental setups are still limited by short coherence times. Our approach points to a new paradigm: by combining measurements with FNO-based extrapolation, one can effectively extend the temporal horizon of quantum experiments beyond their coherence window. This opens a practical route to probing long-time dynamics in regimes that would otherwise be too costly or impossible to probe directly.

\section{Discussion \label{sec:discussion}}
\subsection{Analysis of Computational Complexity}

As we have discussed above, learning observable dynamics from a subset of measurements (Sec.~\ref{sec:obs_to_obs}) and extrapolating  relies on FNO's ability to construct expressive representations of the system's evolution. Such tasks are generally intractable using traditional numerical methods, thus necessitating machine-learning approaches.

However, even for tasks that can be done with conventional numerical methods, such as finding effective Floquet Hamiltonians (Sec.~\ref{sec:Heff}), computing expectation values of local observables for a given Hamiltonian (Sec.~\ref{sec:Ht_to_obs}), and tracking operator evolution (Sec.~\ref{sec:opgrowth}), we will argue that FNO still offers significant practical advantages. In particular, FNO serves as a surrogate model that speeds up computations parametrically. While the cost of exact numerical methods scales exponentially with the system size, we will argue that the runtime and memory cost of FNO approaches only scale polynomially with $L$.

For concreteness, consider a one-dimensional domain $D$ discretized into $N_D$ points. We focus on an FNO with channel width $w$ that retains $k_\text{max}$ frequency modes in its Fourier layers. The primary computational bottleneck of the FNO arises from these Fourier layers~\cite{FNO}. Each layer involves computing a forward and inverse Fourier transform,  $\mathcal{F}$ and $\mathcal{F}^{-1}$, independently across $w$ channels. When implemented using FFT algorithms, the cost of these transforms scales as $\mathcal{O}(wN_D\log N_D)$ per layer. Between $\mathcal{F}$ and $\mathcal{F}^{-1}$, a linear transformation $R_i$ is applied independently to each retained frequency mode, leading to a computational cost that scales as $\mathcal{O}(k_\text{max} w^2)$. Therefore, the computational complexity of each Fourier layer (and thus the FNO) scales as $\mathcal{O}(w N_D \log N_D + k_\text{max} w^2)$.

To understand how the computational complexity scales with the system size, we need to analyze the dependencies of $w$ and $k_\text{max}$ on $L$. First, we argue that $k_\text{max}$ is independent of $L$. The number of retained frequency modes should only depend the behavior of input and output functions in the physical domain (time). When these functions have sharp transitions, jumps, or discontinuities, the FNO generally needs to retain more modes to capture these irregular behaviors. However, physical quantities such as expectation values and Hamiltonian parameters tend to vary smoothly in time, regardless of system size. Increasing $L$ only leads to higher dimensional input and output function spaces (e.g. more Hamiltonian parameters and observables), but does not introduce irregularity in the individual trajectories of these functions. Therefore, when physical quantities are used as inputs and outputs of FNOs, such as in the learning paradigms we have considered, the number of retained Fourier modes $k_\text{max}$ should remain effectively independent of $L$.

Next, we provide a heuristic justification for why the latent space dimension $w$ is expected to scale at most polynomially with $L$. In the examples above, the input and output dimensions $d_i$ and $d_o$, defined in Eq.~\ref{eq:UV}, both scale polynomially in $L$. Therefore, for a given learning task, the operator mapping between the function spaces should also have a rank that scales polynomially in system size. Since the channel width $w$ can be viewed as a proxy for the effective rank of the parameterized neural operator $\mathcal{A}'$~\cite{Kovachki_NeuOp}, we thus expect that $w$ should scale polynomially with $L$. This expectation is supported by empirical observations in our investigation, as well as in previous works~\cite{FNO, Kovachki_NeuOp, scattering_fno}, in which a constant and often small $w$ has been shown to avoid overfitting while being sufficient for accurate modeling in various practical settings.

We emphasize that the computational efficiency of FNO arises from the fact that $d_i$ and $d_o$ scale polynomially in system size, $d_i, d_o = \mathcal{O}(\operatorname{poly}(L))
$. This scaling comes at the expense of necessary approximations that restrict us to a local basis. For example, in Sec.~\ref{sec:Heff}, the inputs and outputs are the weights of $H(t)$ and $H_F(t)$ projected onto the local basis (Eq.~\ref{eq:H_to_vec}). The truncation excludes the exponentially large set of longer-range terms that are present in the exact effective Hamiltonian. Despite this approximation, the FNO achieves remarkable accuracy in predicting stroboscopic dynamics. 

Similarly, in Sec.~\ref{sec:observable}, the FNO can only predict the expectation values of local observables, but not long-range, non-local correlators. However, the local expectation values are typically the ones of greatest experimental relevance and are sufficient for most practical computations. Thus, despite the constraints on locality, the FNO's ability to process local observables makes it a broadly applicable tool effective for many meaningful tasks.

To quantitatively demonstrate the speedup we achieve with FNO, in Table.~\ref{tab:performance} we list the time it takes to complete each task with FNO and exact numerical methods, as well as the relative error of FNO (averaged over each point in time). For fairness of comparison, we have ensured that the exact methods are optimized. We present the time for a single evaluation same discretization ($N=200$ points), each averaged over $100$ runs.  Clearly, even for tasks that exact numerical methods can do, using the FNO achieves a significant speedup, which is more pronounced for the larger system size ($L=8$). Our quantitative benchmarks demonstrate that the FNO offers an alternative method that computes various quantities of interest hundreds of times faster than conventional approaches, with little sacrifice in accuracy. Both methods were performed on the same CPU (AMD Ryzen Threadripper PRO 3975WX). Implementing the FNO on GPUs can speed up forward passes through the neural operator even further.

\begin{table}[]
    \centering
    \begin{tabular}{|c|c|c|c|c|}
     \hline
       Section & $L$ & Exact Time (s) & FNO Time (s) & Relative Error \\ \hline 
     \ref{sec:Heff}  &  8 & 0.728 $\pm$ 0.009 & 0.0050 $\pm$ 0.0002 &  $2.4 \times 10^{-3}$ \\  \hline
     \ref{sec:Ht_to_obs} & 4 &  0.062 $\pm$ 0.002 & 0.0054 $\pm$ 0.0001 & $7.0 \times 10^{-3}$   \\ 
      & 8 &  1.896 $\pm$ 0.043& 0.0041 $\pm$ 0.0001& $2.8 \times 10^{-3}$\\ \hline
      \ref{sec:obs_to_obs} & 4 & N/A & 0.0040 $\pm$ 0.0005 & $8.2 \times 10^{-3}$ \\
       & 8 & N/A & 0.0039 $\pm$ 0.0002 & $4.2 \times 10^{-3}$\\ \hline
     \ref{sec:opgrowth} & 4 & 2.438 $\pm$ 0.014 & 0.017 $\pm$ 0.001 & $1.8 \times 10^{-3}$ \\ \hline 
     \end{tabular}
    \caption{Comparison of run time and accuracy between the exact numerical method and FNO. Each task is repeated for $100$ times on the CPU. 
``N/A'' indicates there is no direct exact numerical analog for mapping between observables; we therefore do not list a baseline time. The FNO achieves remarkable accuracy while providing substantial computational speedup, often by two to three orders of magnitude, across all tasks.}
    \label{tab:performance}
\end{table}

In summary, the total computational complexity of the FNO scales as $\mathcal{O}(\text{poly}(L))$. On the other hand, computing expectation values or effective Hamiltonians exactly requires resources that scale exponentially with system size. While this parametrically large speedup is achieved through approximations that introduce certain limitations, the FNO is a powerful and flexible surrogate model capable of learning from and predicting various experimentally accessible quantities of interest. 

Our results have important and practical implications for leveraging measurements from large-scale quantum computers. As quantum processors continue to scale up in system size and coherence time, the expectation values of local observables, especially for larger system sizes, become more accessible. However, extracting physically meaningful quantities from those data, such as characterizing operator growth and predicting behaviors of observables not measured in experiments, becomes impossible with traditional numerical methods, partly due to the exponential scaling of complexity. The FNO provides a tractable and robust alternative that scales modestly in system size, circumventing the need of knowing or storing full, exponentially large wavefunctions. Our approach enables the use of partial or noisy measurements from quantum hardware to infer global dynamical properties of the system, opening a path toward more scalable characterization and control of quantum many-body systems.

\subsection{Outlook}
There are a few interesting ways in which our investigation can be extended. 

Our study on Floquet quantum systems can be extended to the time-quasi-periodic case. Quasi-periodically driven quantum systems are subjected to multiple mutually incommensurate frequencies and are therefore natural generalizations of Floquet systems. Such systems exhibit exotic non-equilibrium behaviors, such as energy pumping~\cite{pumping1,pumping2,pumping3,pumping4} and novel localization phenomena~\cite{localization1,localization2,localization3}. 

In general, a time-quasi-periodic Hamiltonian under $d$ mutually irrational driving frequencies can be written as:
\begin{equation}
    H(t) = H(\theta_1(t), ..., \theta_d(t))
\end{equation}
with $2\pi$-periodic $\theta_j$, which is parametrically dependent on time via
\begin{equation}
    \theta_j(t) = \omega_j t  + \phi_j.
\end{equation}

Due to the incommensurability of the frequencies, Fourier transform of $H(t)$ lives in a higher-dimensional Fourier space. In particular, we can write:
\begin{equation}
  H(t)= \sum_{\mathbf n \in \mathbb{Z}^d} H_{\mathbf n}\,
e^{-i \mathbf n \cdot \boldsymbol{\theta}(t)},
\end{equation}
where $\boldsymbol{\theta} = (\theta_1(t), \theta_2(t), \dots, \theta_d(t))$.

This higher-dimensional Fourier representation of $H(\boldsymbol{\theta})$ suggests a natural extension of our approach. For example, in the case of two incommensurate drives $\omega_1, \omega_2$, one can employ a 2D Fourier Neural Operator, originally proposed as a surrogate model for solving the Navier-Stokes equation~\cite{FNO}, to learn dynamics of the system in Fourier domain directly. More generally, higher dimensional FNOs could be deployed for systems with multiple incommensurate drives, opening the door to a machine-learning-based framework for modeling quasi-periodically driven quantum systems beyond the reach of standard Floquet techniques.

Another interesting direction for further exploration is the incorporation of physical loss during the training process. In all of our examples above, the loss function is only data-driven i.e. it only measures the discrepancy between the neural operator's predictions with the ground truths. However, additional terms that encode physical constraints can be added. For instance, in Sec.~\ref{sec:Heff}, one can require that the learned effective Hamiltonians $\mathcal{N}(H(t))$ have the same quasi-energy spectrum as $H(t)$ and add a regularization term such as
\begin{equation}
    \mathcal{L}_\text{spectrum} = \lambda \lVert  \sigma(H(t)) -  \sigma(\mathcal{N}(H(t))   \rVert^2
\end{equation}
to the loss function in Eq.~\ref{eq:HF_loss}. Here $\lambda$ is a hyperparameter and $\sigma(\cdot)$ denotes the spectrum of its argument.

Similarly, in Sec.~\ref{sec:opgrowth}, one can directly encode the expected structure of the $c(t)$ matrix into the loss function, Eq.~\ref{eq:opgrowth_loss}. For example, at $t=0$, the system has not evolved at all, so we expect that $c(t=0) = \mathbb{I}_m$, where $m$ is the dimension of the $c$ matrix. We can impose an additional term in the loss function that penalizes the deviation of $c(t=0)$ from identity, such as 
\begin{equation}
    \mathcal{L}_\text{initial condition} = \lambda' \lVert c(t=0)-\mathbb{I}_m \rVert^2
\end{equation}
for some other hyperparameter $\lambda'$. Finally,  we could also encode unitarity into loss function. Training FNO by combining data loss with physic-inspired penalty terms will be an interesting future direction.

Our approach might also be extended to driven, open quantum systems. In such systems, time evolution is governed by a Lindblad master equation~\cite{lindblad1, lindblad2} and is no longer unitary. In regimes where the coupling to the environment is weak, or one is interested in short-time dynamics, time evolution of local observables can still be effectively constrained by locality. In such cases, our approaches for learning local observables and operator growth are expected to remain applicable. Although the inputs will not be strictly periodic in time for open-system dynamics, we have already demonstrated that FNOs can handle non-periodic inputs reliably via padding. We do expect, however, that the temporal window over which predictions remain accurate in an open system should be shorter than in that in the closed-system settings due to couplings to a bath.

Finally, as we have discussed before, it will be interesting to train FNO using measurements from experiments on quantum computers. There have already been nascent works that combine neural networks with experimental measurements~\cite{dual_complexity}. Compared to other machine-learning algorithms and architectures, FNO is robust under noise, sample efficient, agnostic to domain discretization, and naturally suited for time-periodic systems. We leave the intriguing possibility of using an FNO-quantum computer combination to future work.

\section{Conclusion \label{sec:conclusion}}
In this work, we have introduced Fourier Neural Operators (FNOs) as a versatile, accurate, and scalable computational tool for modeling periodically driven quantum systems. Parameterized in Fourier space directly, the FNO serves as a data-driven surrogate that can be trained on short-time data but extrapolate to unseen Hamiltonian parameters, driving frequencies, finer discretizations, and times beyond the training window.

Through three complementary learning paradigms, we demonstrate the broad applicability of FNOs. First, we show that FNOs accurately reconstruct effective Floquet Hamiltonians, even when trained only on coarse temporal data, and generalize across different driving frequencies. Secondly, FNOs can accurately predict expectation values of one- and two-local observables, from either Hamiltonian parameters or partial measurements of the system. We uncover a learnability transition at $k^* \sim L+1$ input observables, beyond which dynamics of all local observables can be accurately learned and predicted. Finally, FNOs are capable of predicting operator-spreading matrices $c(t)$. Using an approximate composition law satisfied by $c(t)$, we are able to extrapolate observable dynamics and track operator growth at times well beyond the training window.

Many of the tasks we have addressed are not just computationally expensive, but effectively impossible, to perform using traditional computational approaches. For example, predicting all local observables from only a subset of measurements and extrapolating dynamics beyond the training window are problems that cannot be solved by exact diagonalization, perturbative expansions, or tensor-network methods. In contrast, FNOs, with their ability to form representations of the underlying physics, transcend these limitations. The ability to infer global dynamical behaviors from limited or partial data is unique to our operator learning framework and suggests a new paradigm for studying non-equilibrium quantum systems, with direct relevance to near-term experiments.

Even for tasks that are possible for conventional numerical approaches, such as computing Floquet Hamiltonians, FNOs still offer significant practical advantages. Once trained, FNOs produce predictions in a single forward pass through the neural operator, bypassing diagonalizations or time evolution, thereby achieving significant speedups during inference. Furthermore, by employing a local basis, we ensure that the number of learnable parameters and therefore computational cost of each learning task only scales polynomially with the system size.

Our results establish the FNO as not only an efficient computational tool, but a novel, data-driven framework for extracting physical predictions from realistic experimental measurements. Our approach not only has efficiency and scalability beyond that of traditional numerical methods, but also offers qualitatively different capabilities of extracting physics from data. Such an integration of FNO-based surrogates with data from NISQ-era platforms provides a powerful pathway toward the scalable characterization, prediction, and control of quantum many-body systems far from equilibrium.

\begin{acknowledgments}
ZQ acknowledges Junkai Dong and Yuan Xue for useful comments on the manuscript.
YP is supported by the US National Science Foundation (NSF) Grants  No.\ PHY-2216774 and No.\ DMR-2406524.
\end{acknowledgments}

\appendix

\section{Hyper-parameters in Fourier Neural Operators \label{app:FNO}}
In this Appendix, we list hyperparameters used for each learning task in this work. Across all tasks, we use $M=3$ Fourier layers and choose $\text{Tanh}$ to be the nonlinear activation function. We also use the AdamW optimizer~\cite{AdamW} with learning rate $0.001$ and weight decay $10^{-6}$. To allow for better convergence and stability, we reduce the learning rate by a factor of $\gamma$ every $n_{\text{decay}}$ steps. We fix $\gamma = 0.8$. 

\begin{figure}
    \centering
    \includegraphics[width=0.85\linewidth]{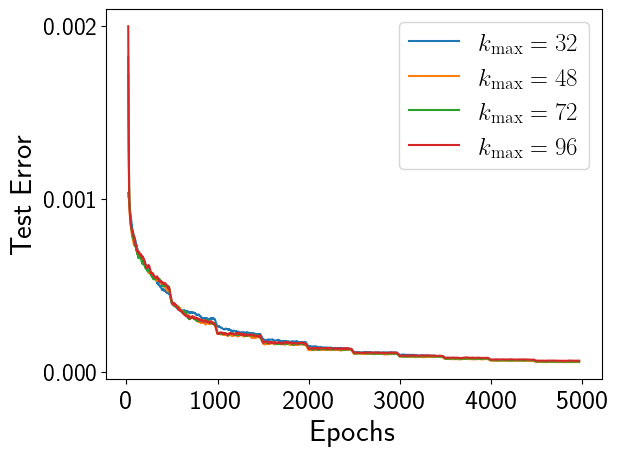}
    \caption{Learning performance of FNO with different numbers of retained frequency modes $k_\text{max}$. Each FNO is trained to approximate the operator mapping from $H(t)$ to $H_F(t)$, as in Eq.~\ref{eq:NNHF}. The test error, given by Eq.~\ref{eq:HF_loss} and window-averaged over 50 epochs, converges for a range of $k_{\max}$, demonstrating that FNO achieves accurate learning of the effective Hamiltonian without requiring a large spectral resolution.}
    \label{fig:kmax_test}
\end{figure}

One of the most important hyper-parameters in an FNO is the number of retained Fourier modes $k_{\text{max}}$. If $k_\text{max}$ is too large, the FNO has less computational efficiency and may risk overfitting; on the other hand, when too few frequency modes are retained, valuable information about the operator may be lost during the truncation. We found that fixing $k_{\max} = 32$ is sufficient for all tasks. As an example, in Fig.~\ref{fig:kmax_test}, we show the test loss during training for various $k_\text{max}$.

The hyper-parameters that change from task to task include: channel width of FNO $w$, number of training samples, number of zeroes added for padding $n_{\text{padding}}$, $n_{\text{decay}}$, and the number of epochs trained. In Table.~\ref{tab:hyperparameters}, we list the hyperparameters used for each learning task.

\begin{table}[]
    \centering
    \begin{tabular}{|c|c|c|c|c|c|c|}
     \hline
       Section & $L$ & $w$ & $\#$ Samples & $n_{\text{padding}}$ & $n_{\text{decay}}$ & Epochs \\ \hline 
     \ref{sec:Heff} & 8  &  256   & 1000 & 0 & 500 & 10000  \\  \hline
    \ref{sec:Ht_to_obs} & 4 & 128  & 600 & 0 & 1000 & 4000   \\ 
    & 8 & 128 & 600 & 0 & 1000 & 4000 \\ \hline
     \ref{sec:obs_to_obs} & 4 & 128 & 4800 & 40 & 500 & 3000 \\ 
     & 8 & 128 & 4800 & 40 & 1000 & 4000\\ \hline
     \ref{sec:opgrowth} & 4 & 256 & 40000 & 40& 10 & 100 \\ \hline 
     \end{tabular}
    \caption{Summary of the hyper-parameters that change between each different learning tasks.}
    \label{tab:hyperparameters}
\end{table}

\section{Proof for Composition Property of $c(t)$ \label{app:ctrelation}}

\newtheorem{proposition}{Proposition}

In this Appendix, we briefly prove the composition relation that is approximately satisfied by $c(t)$, given in Eq.~\ref{eq:c_extrapolation}.

\begin{proposition}
If the set $\{B_j \}$ forms a complete, orthonormal basis of the operator space, the coefficient matrices satisfy the exact relation:
\begin{equation}
    c(t) c(nT) = c(t+nT).
\end{equation}
\end{proposition}

\begin{proof}
We work in the Heisenberg picture, in which operators evolve according to
\begin{equation}
    B_i(t) = U^\dagger(t) B_i U(t),
\end{equation}
where $U(t):=U(t, 0)$ denotes the propagator from $t=0$, defined in Eq.~\ref{eq:evolution}. For time-periodic systems, the unitary time-evolution operator $U(t)$ satisfies the following stroboscopic composition relation:
\begin{equation}
    U(t + nT) = U(t) U(nT) = U(t) U(T)^n.
\end{equation}

We view operators as vectors in operator space, denoted $|A)$. The Hilbert–Schmidt inner product is denoted by
\begin{equation}
    (A|B) = \frac{1}{2^L} \Tr(A^\dagger B).
\end{equation}

Using this notation, the entries of the matrix $c(t)$, defined in Eq.~\ref{eq:cij}, can be written as:
\begin{equation}
    c_{ij}(t) = \left(B_j| B_i(t)\right) = \left( B_j | U^\dagger(t) B_i U(t) \right).
\end{equation}

Now consider the product between two coefficient matrices, $c(t)$ and $c(nT)$:
\begin{equation}
    \begin{aligned}
   [c(t)c(nT)]_{ik} 
   &= \sum_j c_{ij}(t)\,c_{jk}(nT) \\
   &= \sum_j (B_j|B_i(t))\,(B_k|B_j(nT)) \\
   &= \sum_j (U^\dagger(t) B_i U(t) |B_j) (B_j| U(nT) B_k U^\dagger(nT) ).
\end{aligned}
\label{eq:cc_entry}
\end{equation}

If the set $\{B_j \}$ forms a complete, orthonormal basis of the operator space, then it would satisfy the completeness relation
\begin{equation}
    \sum_j |B_j)(B_j| = \mathbb{I},
\end{equation}
using which we could simplify Eq.~\ref{eq:cc_entry} to:
\begin{equation}
\begin{aligned}
      [c(t)c(nT)]_{ik} &=  (U^\dagger(t) B_i U(t) | U(nT) B_k U^\dagger(nT) ) \\
      &= (B_k | U^\dagger(nT) U^\dagger(t) B_i U(t) U(nT)) \\
      &= (B_k | U^\dagger (t+nT) B_i U(t+nT))  \\
      &= [c(t+nT)]_{ik},
      \end{aligned}
\end{equation}
yielding an \textit{exact} relation, $c(t) c(nT) = c(t+nT)$, in the \textit{complete} basis.
\end{proof}

In practice, however, we restrict to a truncated local basis (the set of one- and two-local operators), so the completeness relation is instead replaced by a projector onto the local subspace. The exact relation for the full basis therefore becomes:
\begin{equation}
    c(t) c(nT) = c(t + nT) + r(n, t),
\end{equation}
where $r(t, n)$ is a residual term that quantifies how much weight leaks out of the local basis. At short times, or in the high-frequency limit, operator spreading is weak ($r(n, t) \ll c(t+nT)$), and we can therefore write the following approximate relation:
\begin{equation}
    c(t)c(nT) \simeq c(t+nT).
\end{equation}

\section{Floquet-Magnus Expansion\label{app:magnus}}
The Floquet-Magnus expansion is a common technique for computing the Floquet Hamiltonian in the high frequency limit~\cite{magnus, magnus1}. The method approximates the Floquet Hamiltonian as a perturbative expansion in powers of $1/\omega$, with the two leading terms being:
\begin{align}
H_F^{(0)} &= \frac{1}{T} \int_0^T H(t) \, dt, \\
H_F^{(1)} &= \frac{1}{2iT} \int_0^T dt_1 \int_0^{t_1} dt_2 \, [H(t_1), H(t_2)].
\end{align}

While the Magnus expansion is not guaranteed to converge at intermediate to low frequencies, it does capture the qualitative features of the driven dynamics when $\omega$ is the highest energy scale of the system, $\omega \gg J, A, \eta$. In the main text, we compare the performance of our FNO's learned effective Hamiltonian against the Floquet Hamiltonian obtained from second-order Magnus expansion.

\section{Padding for Non-Periodic Inputs \label{app:padding}}

\begin{figure}
    \centering
    \includegraphics[width=1\linewidth]{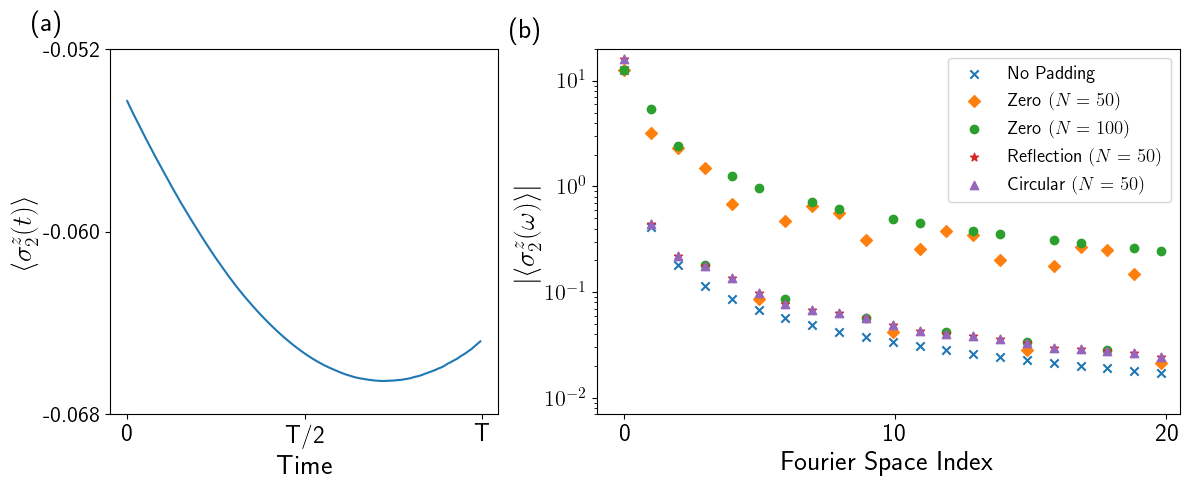}
    \caption{(a) Dynamics of a typical observable, $\sigma_2^z(t)$, over one period. (b) Weights of the Fourier components after applying different methods of padding. While circular and reflection padding smooth out the jump slightly, zero padding produces the smallest spectral leakage in our tests. }
    \label{fig:padding}
\end{figure}

Traditional Fourier methods assume periodic inputs, which can be expressed as a Fourier series. The Fourier Neural Operator, on the other hand, does not have this limitation due to the linear transform $b_i(x)$ in the Fourier layers~\cite{FNO}. FNO has been shown to be effective for various problems even when the input functions are not strictly periodic~\cite{FNO, FNO_padding, FNO_padding2, FNO_padding3, FNO_padding4, nonperiodic, nonperiodic1, nonperiodic2}.

However, for non-periodic inputs, it is still important to reduce the numerical artifacts associated with the discontinuity as the input wraps around the domain. Common artifacts from the mismatch between $f(0)$ and $f(N)$, with $N$ being the length of the input sequence, include spectral leakage and the Gibbs phenomenon~\cite{DFT_spectralleakage,Gibbs_phenomenon}. Padding the inputs is one of the most common and effective ways of smoothing out the jump when non-periodic functions are inputted to FNO~\cite{FNO, FNO_padding2, FNO_padding3, FNO_padding4}.

There are three common ways to pad input functions: zero padding, reflect padding, and circular padding. For concreteness, consider the discretized domain $D_j = \{t_0, t_1, ..., t_{N-1} \}$, and a non-periodic function $f$ is evaluated only at these points. Zero padding appends zeroes after the last value $f(t_{N-1})$. Reflect padding extends the input by adding interior values in reverse order: $f(t_{N-2}), f(t_{N-3}),...$ are added after $f(t_{N-1})$. Finally, circular padding adds value from the other end after the sequence. In this case, $f(t_0), f(t_1), ...$ are appended after $f(t_{N-1})$. Regardless of the method, the padded output is truncated back to the original length after the Fourier transform.

To investigate the efficacy of the three methods, we take the expectation value of a typical observable $\left< \sigma_2^z(t) \right>$ over one period as the input (see Fig.~\ref{fig:padding}(a)). As the domain $D$ wraps around, the expectation value is smooth but non-periodic, $\left< \sigma_2^z(t = 0) \right> \neq \left< \sigma_2^z(t = T) \right>$. As a result, its Fourier transform has a peak in the early harmonics. The goal of padding is to smooth out this abrupt jump.

In Fig.~\ref{fig:padding}(b), we show the modulus of weights of the Fourier transform, $\left| \left< O_i(\omega) \right> \right| = \left| \int_0^T \left< O_i(t) \right> e^{i\omega t} dt \right|$ for each method of padding. Clearly, zero padding is the method that allows for the smoothest and slowest decay of the weights in Fourier space, enhancing the expressivity of FNO. We will use zero padding for non-periodic inputs throughout this work.

\section{FNO Performance for Entangled Initial States}
In the main text, we have focused on cases where the initial state of the system is a product state. However, time-evolution under the disordered TFIM inevitably generates entanglement between the sites. To demonstrate that our approach remains valid in the presence of entanglement, in this Appendix, we consider the two tasks in Sec.~\ref{sec:observable}, namely learning local observables from Hamiltonian parameters and learning two-local correlators from one-local observables, for initial states with varying levels of entanglement.

To generate entangled initial states, we will first randomly generate product states, with each sites being a random two-level state sampled from the Haar measure. We will then apply the following gate to neighboring sites:
\begin{equation}
    C(\theta) = \begin{pmatrix}
1 & 0 & 0 & 0 \\
0 & 1 & 0 & 0 \\
0 & 0 & 1 & 0 \\
0 & 0 & 0 & e^{i\theta}
\end{pmatrix},
\end{equation}
and use 
\begin{equation}
    \ket{\psi(\theta)} = \bigotimes_{i} C_{i, i+1}(\theta) \ket{\psi_0}
\end{equation}
as our entangled initial state. The tunable parameter $\theta$ controls the amount of initial entanglement: $\ket{\psi(\theta=0)}$ reduces to a product state, while $\ket{\psi(\theta = \pi)}$ has the maximum entanglement under applications of this gate to neighboring sites.

Specifically, we will consider three values of $\theta$: $\theta = \pi/10, \pi/3, \text{and }\pi$, corresponding to initial states with varying degrees of entanglement. For each $\theta$, we generate random Hamiltonians and numerically simulate expectation values of local observables to use for training.

\begin{figure}
    \centering
    \includegraphics[width=1\linewidth]{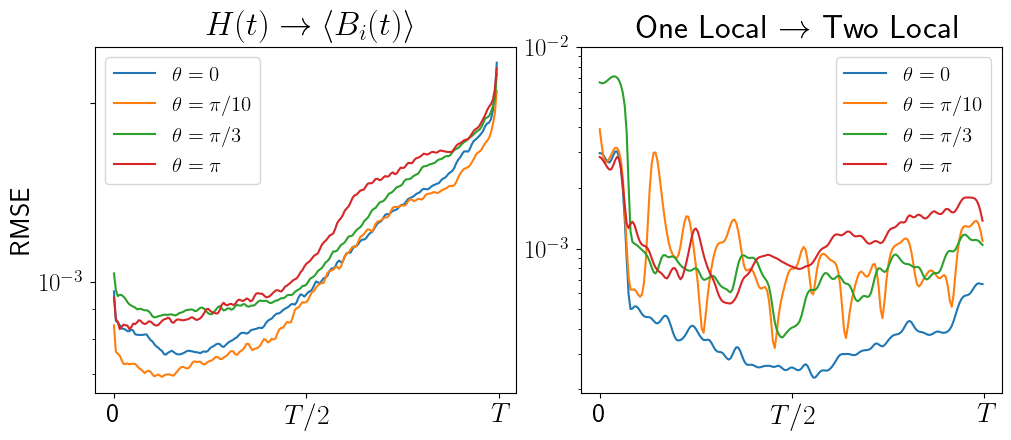}
    \caption{Performance for initial states with various levels of entanglement, as controlled by the parameter $\theta$. The Root-Mean Squared Error (RMSE), defined in Eq.~\ref{eq:rmse}, is largely independent of the level of initial entanglement.}
    \label{fig:entanglement}
\end{figure}

With the data prepared, next we train an FNO to learn the mappings from $H(t)$ to $\{B_i(t)\}$ and from one-local observables to two-local observables, similar to Secs.~\ref{sec:Ht_to_obs} and ~\ref{sec:obs_to_obs}. To make sure the comparison is fair, we have used the \textit{same} hyper-parameters as ones used in their respective sections. As shown in Fig.~\ref{fig:entanglement}, while the FNO seems to perform slightly better for the unentangled initial state, the RMSE averaged over all predicted observables is mostly insensitive to the level of entanglement and remains on the order of $10^{-3}$.

\section{Learning Local Observables for Larger System Size~\label{app:L14}}
In the main text, we have presented numerical results from system sizes of $L=8$ and $L=4$. While we have discussed the scalability of FNO with system size in Sec.~\ref{sec:discussion}, in this Appendix we concretely demonstrate the applicability of FNOs to larger system sizes. We still consider the periodically driven TFIM with spatiotemporal disorder (Eq.~\ref{eq:TFIM}), but with a larger system size $L=14$. We show that FNOs are not only capable of predicting local observables accurately, but remain orders of magnitude faster than conventional numerical methods.

Due to the exponential growth of the Hilbert space dimension, exact dense-matrix time evolution quickly becomes prohibitively expensive in both memory and runtime for larger system sizes. Therefore, we switch to Krylov–subspace–based time evolution~\cite{ParkLight1986, Saad1992} with a sparse representation of the Hamiltonian, storing only non-zero elements. Instead of exponentiating the full matrix, we approximate the action of the short-time propagator $\exp(-i H(t) \Delta t)$ directly on the state vector using a Krylov algorithm for the matrix exponential~\cite{AlMohyHigham2011}. We further optimized the measurements of local observables by using the sparsity of the Pauli operators, computing expectation values via direct vector dot products rather than full matrix multiplication. Finally, the simulation is parallelized across multiple CPU cores to ensure that collecting thousands of time traces is possible in a practical amount of time.

\begin{figure}[t!]
    \centering
    \includegraphics[width=0.75\linewidth]{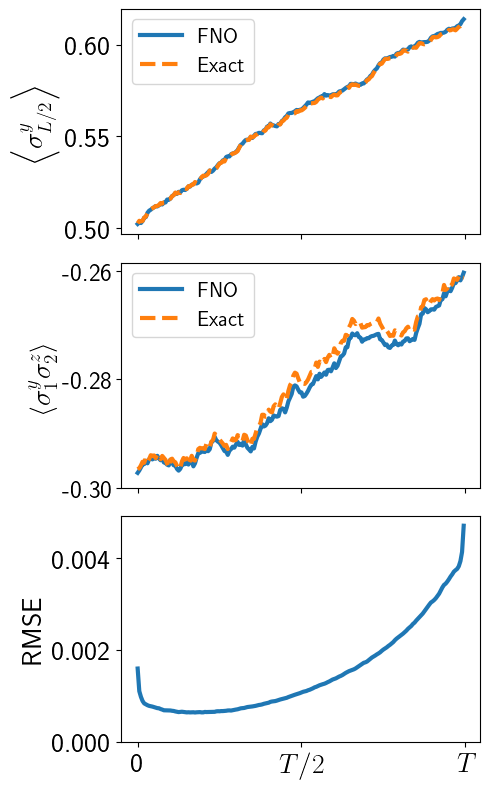}
    \caption{FNO performance for $L=14$. All one- and two-local observables can be learned to great accuracy. The RMSE, averaged over all observables, remains on the order of $10^{-3}$, on the same order as $L=8$.  }
    \label{fig:L14}
\end{figure}

With data from $L=14$, we train an FNO that learns the mapping from $H(t)$ to $\{ \left< B_i(t) \right> \}$. This task is chosen to allow for a direct comparison in runtime with other numerical techniques. In Fig.~\ref{fig:L14}, we show the performance of FNO on predicting local observables for unseen Hamiltonian parameters. Both one-local and two-local correlators can be learned with great precision; the Root-Mean Squared Error (Eq.~\ref{eq:rmse}), averaged over all one- and two-local observables, remains on the order of $10^{-3}$ during the entire time interval.

Averaged over 100 runs on the CPU, the FNO takes an average of 53.6 ms at inference, roughly ten times the inference time for $L=8$; see Table.~\ref{tab:performance}. By contrast, even the Krylov-space-based method, which is already \textit{significantly} faster than exact time-evolution, takes an average of $13.1$ s per run. The FNO approach remains orders of magnitude faster than conventional numerical solvers.

We summarize how the FNO's run time and relative error scale with system size in Table.~\ref{tab:L_scaling}. We compare the performance of the same task, namely learning local observables $\left< O_i(t) \right>$ from Hamiltonian parameters $H(t)$. The dynamics run over $8$ periods for $L=4$ and one period for $L=8$ and $L=14$. The ``exact'' numerical method refers to stepwise Trotterized evolution for $L=4$ and $L=8$, while it refers to Krylov-subspace time evolution~\cite{ParkLight1986, Saad1992} for $L=14$. In all cases, the relative error is averaged over all local observables and all points in time. The FNO's inference time is orders of magnitude faster than traditional numerical methods, while capturing local dynamics highly accurately.

\begin{table}[]
    \centering
    \begin{tabular}{|c|c|c|c|}
     \hline
        $L$ & Exact Time (s) & FNO Run Time (s) & Relative Error \\ \hline 
     4 & $0.062 \pm 0.002$ & $0.0054 \pm 0.0001$ & $7.0 \times 10^{-3}$ \\ \hline
     8 & $1.896 \pm 0.043$ & $0.0041 \pm 0.0001$ & $2.8 \times 10^{-3}$ \\ \hline
     14 & $13.1 \pm 1.3$ & $0.0536 \pm 0.0012 $ & $1.2 \times 10^{-2}$ \\ \hline
     \end{tabular}
    \caption{Summary of the scaling of run time and relative error with system size, for $L=4, 8, 14$, for the task of predicting local observables from Hamiltonian parameters. All run time computations are averaged over $100$ runs. The FNO's run time remains faster by orders of magnitude, while capturing the dynamics accurately.}
    \label{tab:L_scaling}
\end{table}

\section{FNO Performance for 2D Triangular Lattice\label{app:triangular}}
To demonstrate the versatility of our framework beyond one-dimensional chains, we apply the FNO method to a two-dimensional triangular lattice with dimensions $3 \times 4$, consisting of $L=12$ spins. The Hamiltonian we consider is still the time-dependent two-dimensional transverse field Ising model, with spatio-temporal disorder:
\begin{equation}
    H(t) = \sum_{\left<ij \right>} J_i X_i X_j + \sum_{i=1}^L A_i \cos(\omega t) Z_i + h_x(i, t) X_i,
\end{equation}
where $\left< ij \right>$ denotes bonds between nearest neighbors and $L$ is the total number of sites. Here $J_i$, $A_i$, and $h_x(i, t)$ are still sampled uniformly and independently as described in Sec.~\ref{sec:model}. Periodic boundary conditions in both directions are assumed.

\begin{figure}[t!]
    \centering
\includegraphics[width=0.8\linewidth]{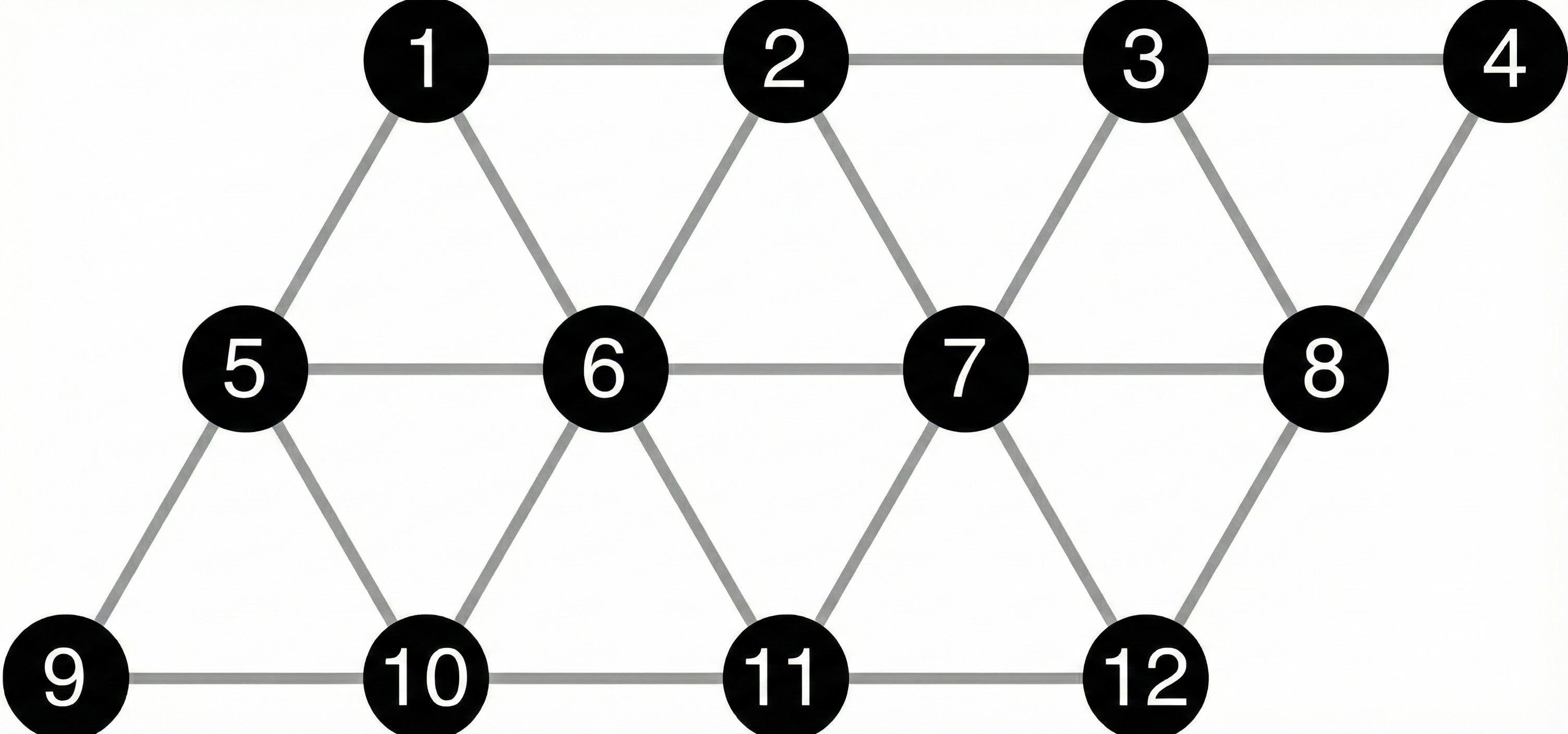}
\caption{Illustration of the triangular lattice, as well as the numbering scheme, that we use in Appendix.~\ref{app:triangular}. Periodic boundary condition is imposed in both $x$ and $y$ directions.}
    \label{fig:triangular}
\end{figure}

While the physical connectivity of this model is two-dimensional, we treat the input data as an effectively one-dimensional sequence to maintain compatibility with the FNO architecture in Sec.~\ref{sec:fno}, which only performs convolution in the time domain. In particular, we perform a flattening (or serialization) operation. We assign an index $i \in \{1, 2, \dots, L \}$ to each site $(x, y)$ in the 2D lattice, as illustrated in Fig.~\ref{fig:triangular}.

Under this mapping, the geometrically local couplings in 2D transform into a set of 1D couplings with varying ranges. For example, for the lattice we consider, a nearest-neighbor coupling in the $y$-direction connects indices $i$ and $i+4$ (mod $L$) in the 1D sequence. Consequently, the effective 1D Hamiltonian contains both nearest-neighbor terms and longer-range interactions.

\begin{figure}[t!]
    \centering
    \includegraphics[width=0.65\linewidth]{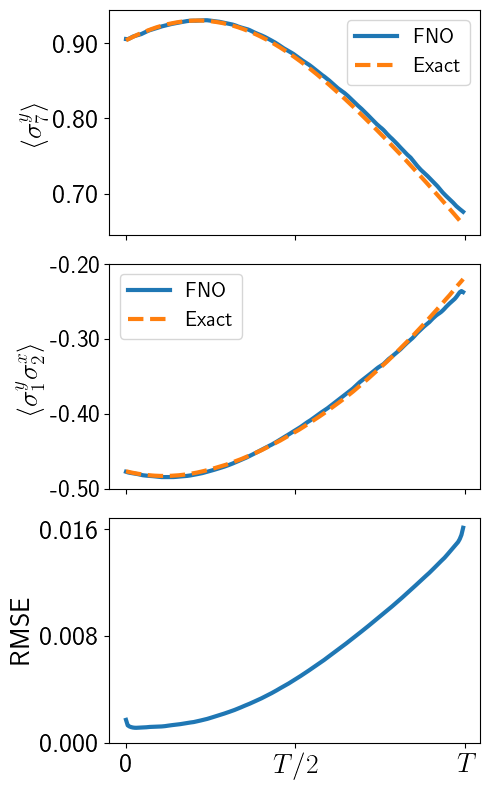}
    \caption{FNO performance for learning local observables the TFIM on a triangular lattice. All one- and two-local observables can be predicted by the FNO, for an unseen Hamiltonian $H_{\text{2D}}(t)$. The RMSE is a few times larger compared to the 1D TFIM, but remains on the order of $10^{-2}$.}
    \label{fig:FNO_triangular}
\end{figure}

As demonstrated in Fig.~\ref{fig:FNO_triangular}, for the 2D triangular lattice we consider, the FNO learns the mapping $H(t) \rightarrow \{ \left< B_i(t) \right> \}$ with reasonable accuracy. While the RMSE averaged over all observables is a few times larger than that in 1D, this example still highlights the predictive power of FNO, even in the presence of geometric frustration.

We have two remarks on the extension of FNOs to 2D. First, different from 1D lattices, sites on a 2D lattices can have more than two neighbors. As a consequence, the number of two-local observables can be significantly larger than that in 1D models. For the triangular lattice with PBC, each site is connected to 6 neighbors. There are, therefore, $3L$ unique bonds and $27L$ two-local observables in the system, while the number of one-local (onsite) observables remains at $3L$. The input and output dimensions of FNO applied to a 2D system are therefore larger than those applied in 1D models with the same system size.

Second, we stress that the FNO architecture itself requires \textit{no modification} when adopted to 2D lattices. The 1D Fourier layers perform global convolutions in the frequency domain only. Because the Fourier transform mixes information from all spatial positions simultaneously, the FNO does not rely on the local sparsity of the graph.

%

\end{document}